\documentclass[structabstract]{aa}
\usepackage{graphicx}
\usepackage{natbib} \bibpunct{(}{)}{;}{a}{}{,}
\usepackage{color}
\usepackage{rotating}
\usepackage{txfonts}

\def\changedA{}
\def\changedB{}
\def\changedC{}

\def\changed{}
 
 \newcommand{\msunpyr}{\,M_\odot\,\mbox{yr}^{-1}}

\newcommand{\kms}{\ifmmode{\,\mbox{km}\,\mbox{s}^{-1}}\else{km/s}\fi}
\newcommand{\msun}{\ifmmode M_{\odot} \else M$_{\odot}$\fi}
\newcommand{\rsun}{\ifmmode R_{\odot} \else R$_{\odot}$\fi}
\newcommand{\lsun}{\ifmmode L_{\odot} \else L$_{\odot}$\fi}
\newcommand{\zsun}{\ifmmode Z_{\odot} \else $Z_{\odot}$\fi}
\newcommand{\velo}{\ifmmode\varv\else$\varv$\fi}
\newcommand{\vinf}{\ifmmode\velo_\infty\else$\velo_\infty$\fi}
\begin{document} 
 
\title{The Eddington factor as the key to understand the winds of the most massive stars}
\subtitle{Evidence for a $\Gamma$-dependence of Wolf-Rayet type mass loss}

\titlerunning{\changedA The Eddington factor as the key to understand the winds of the most massive stars}
 
\author{G.\ Gr\"{a}fener\inst{\ref{inst1}}
\and    J.S.\ Vink\inst{\ref{inst1}}
\and    A.\ de Koter\inst{\ref{inst2},\ref{inst4}}
\and    N.\ Langer\inst{\ref{inst3},\ref{inst4}}
}
 
\institute{Armagh Observatory, College Hill, Armagh BT61\,9DG, United Kingdom\label{inst1}
\and Astronomical Institute ``Anton Pannekoek'', University of Amsterdam, Science Park XH, Amsterdam, The Netherlands\label{inst2}
\and Argelander-Institut f{\"u}r Astronomie der Universit{\"a}t Bonn, Auf dem H{\"u}gel 71, 53121 Bonn, Germany\label{inst3}
\and Astronomical Institute, Utrecht University, Princetonplein 5, 3584 CC Utrecht, The Netherlands\label{inst4}
}

 
\date{Received ; Accepted}

\abstract{\changedA The most massive stars {{\changedB are thought to
      be}} hydrogen-rich Wolf-Rayet stars of late spectral subtype (in
  the following WNh stars). The emission-line spectra of these stars
  are indicative of strong mass loss. In previous theoretical studies
  this enhanced mass loss has been attributed to their proximity to
  the Eddington limit.}  {We investigate observed trends in the
  mass-loss properties of {{\changedB such}} young, very massive stars
  to examine a potential {{\changedB $\Gamma$-dependence, i.e., with
      respect to the classical Eddington factor $\Gamma_{\rm e}$.}}
  Based on different mass estimates, we gain information about the
  evolutionary status of these objects.}  {We derive theoretical
  mass--luminosity relations for very massive stars, based on a large
  grid of stellar structure models. Using these relations, we estimate
  Eddington factors ($\Gamma_{\rm e}$) for a sample of stars, under
  different assumptions of their evolutionary status. We evaluate the
  resulting mass-loss relations, and compare them with theoretical
  predictions.}  {We find observational evidence that the mass loss in
  the WR regime is dominated by the Eddington parameter $\Gamma_{\rm
    e}$, {{\changedA which has important consequences for the way we
      understand Wolf-Rayet stars and their mass loss.  In addition,
      we derive}} wind masses that support the picture that the WNh
  stars in young stellar clusters are very massive, hydrogen-burning
  stars.}  {Our findings suggest that the proximity to the Eddington
  limit is the physical reason for the onset of Wolf-Rayet type mass
  loss. This means that, e.g.\ in stellar evolution models, the
  Wolf-Rayet stage should be identified by large Eddington
  parameters, instead of {\changedA a helium-enriched surface
    composition}.  The latter is most likely only a consequence of
  strong mass loss, {\changedA in combination with internal mixing}.
  For very massive stars, the enhanced $\Gamma$-dependent mass loss is
  responsible for the formation of late WNh subtypes {{\changedB with
      high hydrogen surface abundances, partly close to solar.}}
  Because mass loss dominates the evolution of very massive stars, we
  expect a strong impact of this effect on their end products, in
  particular on the potential formation of black holes, and Gamma-Ray
  Bursts, as well as the observed upper mass limit of stars.}
\maketitle

\section{Introduction} 
\label{sec:intro} 

Wolf-Rayet (WR) type mass loss fundamentally affects the evolution,
the final fate, and the chemical yields of massive stars. The amount
of mass loss in the WR phase {\changedA predominantly} decides whether
a star ends its life as neutron star, or black hole \citep{heg1:03}.
In particular, WR-type mass loss at low metallicities ($Z$) is
expected to be of paramount importance for the chemical enrichment of
the early universe \citep[][]{mey1:06,chi1:06}, and the formation of
long-duration gamma ray bursts \citep[long GRBs,][]{yoo1:05,woo1:06}.

The nature of WR-type mass loss is however still poorly understood.
Stellar evolution models mostly rely on empirical mass-loss relations
\citep[e.g.][]{nug1:00,ham1:06}, with the WR phase
identified on the basis of
observed WR surface abundances in our {\changedB galaxy}. Clearly, for
the modeling of stellar populations that cannot be observed locally, a
more physical approach would be desirable.

In the present work we elaborate on such an approach, namely a
mass-loss relation for WR stars that chiefly depends on the Eddington
factor $\Gamma_{\rm e}$ (Eq.\,\ref{eq:gammae}).  Such relations have
been predicted {\changedA for {\changedC very massive stars} close to the
  Eddington limit \citep{gra1:08,vin1:11}, and for LBVs
  \citep{vin1:02}.}  Notably, the proximity to the Eddington limit
provides a natural explanation for the occurrence of the WR
phenomenon.  Large Eddington factors can be reached, on the one hand,
by {\changedB very massive stars} on the main sequence because of their
extremely high luminosities, and on the other hand, by less massive
evolved (He-burning) stars due to the enhanced mean molecular weight
in their cores. The occurrence of WR-type mass loss for young,
luminous, hydrogen-rich stars (typically late WNh subtypes), and
evolved, hydrogen-free WR stars (nitrogen-rich WN, and carbon-rich WC
subtypes) can thus be explained in the same way.

{\changedA The main goal of this paper is to} confront the
theoretically predicted concept of $\Gamma$-dependent mass loss with
the observed mass-loss properties of very {\changedA massive stars. We
  use the results of a study of the most massive stars in the Arches
  cluster near the Galactic centre (GC), by \citet{mar1:08}. The
  Arches cluster is a young star forming region rich in very massive
  stars, including many O and Of+ supergiants, and luminous WNh stars.
  It forms an ideal testbed for our present study.}

{\changedA Our paper is organized in the following way. In
  Sect.\,\ref{sec:obs} we briefly describe the properties of luminous
  WNh stars.  Sect.\,\ref{sec:pred} recaps our theoretical knowledge
  of the mass-loss properties of these objects.  To study the
  dependence of mass loss on $\Gamma_{\rm e}$ requires mass estimates
  for the sample stars. In Sect.\,\ref{sec:masses} we provide mass
  versus luminosity relations for chemically-homogeneous stars, that
  may be used to derive the Eddington factor for the Arches stars from
  their observed luminosities, and surface abundances. In
  Sect.\,\ref{sec:arches}, we study the dependence of the mass-loss
  rates on stellar properties, including $\Gamma_{\rm e}$, and compare
  these dependencies with theoretical predictions.  The main intention
  in this section is to perform a {\em qualitative} comparison. At the
  present time a quantitative comparison may be affected by systematic
  uncertainties.  Our findings are discussed in
  Sect.\,\ref{sec:discussion}, and conclusions are drawn in
  Sect.\,\ref{sec:concl}.}

\section{Physical properties of the most massive stars}
\label{sec:obs}

The most massive stars with direct mass estimates are found in binary
systems that contain luminous, hydrogen-rich WNh stars. The highest
masses lie in the range of 70--$120\,M_\odot$
\citep{rau1:96,sch1:99,rau1:04,bon1:04,sch1:08,sch1:09}.  Furthermore,
spectral analyses of (putatively) single WNh stars imply very high
masses.  Typical luminosities lie in the range of $10^6\,L_\odot$ or
higher, implying that most WNh stars are very massive stars in the
phase of core H-burning
\citep{cro1:95,cro1:97,cro1:98,dek1:97,ham1:06,mar1:08,mar1:09,cro1:10}.

Luminous WNh stars are preferentially found in the centers of massive,
young stellar clusters with ages of only a few Myr. {\changedA This,
  combined with the fact that their surfaces still show ample amounts
  of hydrogen, suggests that they are still in their phase of core
  H-burning.}  Well-known examples are the Arches cluster close to the
Galactic centre, the young galactic cluster NGC\,3603, and R\,136, the
central cluster of the star-forming region 30\,Dor in the {\changedA
  large Magellanic cloud (LMC)}.  \citet{cro1:10} recently determined
a luminosity of $10^{6.94}\,L_\odot$ for the brightest star in R\,136,
which would correspond to a single star with a present-day mass of
$265\,M_\odot$.

The fact that the most massive stars appear as Wolf-Rayet spectral
types can be explained {\changedA as a result of} their proximity to
the Eddington limit.  According to \citet{gra1:08}, the increased
density scale height close to the Eddington limit leads to the
formation of strong winds with large optical depths. Because ionizing
photons are efficiently absorbed within these winds, recombination
sets in. This shows up in the form of strong WR emission lines that
originate from the subsequent recombination cascades.

In the following we characterize the proximity to the Eddington limit
by the `classical' Eddington factor
\begin{equation}
  \label{eq:gammae}
  \Gamma_{\rm e}=\chi_{\rm e}L/(4\pi c\, G M),
\end{equation}
with the electron scattering opacity $\chi_{\rm e}$\footnote{\changedB
  Note that the opacity $\chi$ is a mass absorption coefficient, i.e.,
  in the CGS system it is measured in ${\rm cm}^2/{\rm g}$.}.  Because
H, and He are completely ionized in the inner regions of hot star
atmospheres, $\Gamma_{\rm e}$ is nearly constant with radius, and
depends only on the stellar parameters $M$, $L$, and, via the mass
absorption coefficient $\chi_{\rm e}$, on the hydrogen mass fraction
$X_{\rm H}^s$ at the stellar surface (Eq.\,\ref{eq:gamma}).

{\changedA There exists a significant additional contribution to the
  total mean opacity $\chi(r)$, due to metal lines (chiefly Fe) and
  continua. For the {\em physical},} radius dependent Eddington factor
$\Gamma(r)$, including all opacities, we thus have
\begin{equation}
 \label{eq:gammar}
  \Gamma(r) = \chi(r) L/(4\pi c\, G M) > \Gamma_{\rm e}.
\end{equation}
The size of the shift between $\Gamma$ and $\Gamma_{\rm e}$ depends on
the detailed metal abundances, and on the ionization structure of the
atmosphere, and thus on $T_\star$, and $Z$.  E.g., for solar
metallicity models, \citet[][]{gra1:08} find that $\Gamma$ approaches
unity in deep atmospheric layers, {\changedB already} for $\Gamma_{\rm
  e} \approx 0.5$.  The onset of WR-type mass loss thus occurs for
{\changedB $\Gamma_{\rm e}<1$, i.e., for Eddington factors that are
  smaller, but still of the order of one.}

\section{Mass-loss predictions for very massive stars}
\label{sec:pred}

Mass-loss predictions for very massive stars have been performed by
\citet{gra1:08}, and \citet{vin1:11}. Despite the rather different
modeling approaches, these works agree on the dominant role of the
Eddington factor $\Gamma_{\rm e}$ for the mass-loss properties of very
massive stars.

The mass-loss {\changedA relation by \citet{gra1:08} is based on}
advanced stellar atmosphere models that incorporate non-LTE line
blanketing, wind clumping, and an exact numerical solution of the
hydrodynamic equations \citep{gra1:05}. {\changedC The models include
  complex model atoms of H, He, C, N, O, Si, and the Fe-group, which
  should be sufficient to describe the largest part of the radiative
  wind acceleration. We note however that the lack of intermediate
  elements (Ne--Ca) could potentially lead to an under-estimation of
  the mass-loss rates.}

Apart from the dominant role of $\Gamma_{\rm e}$, \citeauthor{gra1:08}
find a strong dependence on $T_\star$, and a strong $Z$-dependence.
Their mass-loss prescription has the form
\begin{equation}
  \label{eq:massloss1}
  \begin{array}{l}
    \log(\dot{M}) = -3.763\\
 \rule{0cm}{2.4ex}\hspace{1.02cm} 
    +\, \beta \cdot \log\left(\Gamma_{\rm e} - \Gamma_0\right)
        - 3.5\cdot\left(\log (T_\star/{\rm K}) - 4.65\right)\\
 \rule{0cm}{2.4ex}\hspace{1.02cm} 
    +\, 0.42\cdot\left(\log (L/L_\odot)-6.3\right)
      - 0.45\cdot\left(X_{\rm H} - 0.4\right),\\
  \end{array}
\end{equation} 
with $\dot{M}$ in $\msunpyr$, and the $Z$-dependent parameters
$\beta$, and $\Gamma_0$ given by
\begin{eqnarray}
  \label{eq:fit2}
  \beta(Z) &=& 1.727 + 0.250 \cdot \log(Z/Z_\odot), \\
  \label{eq:fit3}
  \Gamma_0(Z) &=& 0.326 - 0.301 \cdot \log(Z/Z_\odot) - 0.045 \cdot \log(Z/Z_\odot)^2.
\end{eqnarray}
Note that the stellar temperature $T_\star$ denotes the effective
core temperature as defined, e.g., in \citet{gra1:02}.

\citet{vin1:11} compute mass-loss rates {\changedA for very massive
  stars} using a Monte Carlo approach with a parameterized solution of
the hydrodynamic equations \citep{mul1:08}.  They confirm the dominant
role of $\Gamma_{\rm e}$, but find a weak temperature dependence in
the range of 30--50\,kK.  Notably, \citeauthor{vin1:11} resolve the
transition between classical OB star mass loss, with a relatively weak
dependence on $\Gamma_{\rm e}$ (Eq.\,\ref{eq:vink1}), and WR-type mass
loss with a much steeper dependence (Eq.\,\ref{eq:vink2}). However,
\citeauthor{vin1:11} note that the precise value of $\Gamma_{\rm e}$
where this transition occurs in their models, might be too high. The
likely reason is that the shift between the Eddington factor
$\Gamma(r)$, and $\Gamma_{\rm e}$ (cf.\ Eq.\,\ref{eq:gammar}) is
under-estimated in their models (see also the discussion in
Sect.\,\ref{sec:disc_mdot}).

For a solar composition, and an effective temperature of 50\,kK
\citeauthor{vin1:11} give relations of the form
\begin{equation}
\label{eq:vink1}
    \log(\dot{M}) \propto
           1.52 \log(\Gamma_{\rm e})
          + 0.68 \log(L/L_\odot)
          \;\;\;\;\;\;\; \Gamma_{\rm e}<0.7
\end{equation} 
\begin{equation}
\label{eq:vink2}
    \log(\dot{M}) \propto
           3.99 \log(\Gamma_{\rm e})
          + 0.78 \log(L/L_\odot)
          \;\;\;\;\;\;\; \Gamma_{\rm e}>0.7,
\end{equation} 
with $\dot{M}$ in $\msunpyr$.

For the Eddington factor $ \Gamma_{\rm e}$ both works adopt the value
for a fully ionized plasma, which is given by
\begin{equation}
  \label{eq:gamma}
  \log(\Gamma_{\rm e}) = -4.813 + \log(1+X_{\rm H}^s)
  + \log(L/L_\odot)
  - \log(M/M_\odot).
\end{equation} 
In this form $\Gamma_{\rm e}$ only depends on the stellar parameters
$M$, $L$, and, via $\chi_{\rm e}$, on the hydrogen mass fraction
$X_{\rm H}^s$ at the stellar surface (cf.\ Eq.\,\ref{eq:gammae}).  

The strong sensitivity to the Eddington factor in both mass-loss
relations\footnote{\changedA The steepness of the $\Gamma_{\rm
    e}$-dependence in Eq.\,\ref{eq:massloss1} is mainly due to the
  fact that $\log\left(\Gamma_{\rm e} - \Gamma_0\right)$ is evaluated
  for $\Gamma_{\rm e} - \Gamma_0$ close to zero.} offers the potential
to provide very precise estimates of $\Gamma_{\rm e}$ for specific
objects.  If $\dot{M}$, $L$, $X_{\rm H}^s$, and $Z$ are known from
spectral analyses, it is thus possible to obtain very precise mass
estimates, within the systematic errors of the adopted mass-loss
relation.

A major goal of the present work is to calibrate the underlying
mass-loss relation by a comparison of such `wind masses' with
predicted masses from stellar structure computations. If the important
dependencies on $\Gamma_{\rm e}$, $T_\star$, and $Z$ are backed up by
observations, the mass-loss relations may not only serve as input for
stellar evolution computations, but can also provide an important
diagnostic tool to examine the present masses, and thus the
evolutionary status of observed stars.

\section{Mass--luminosity relations for very massive stars}
\label{sec:masses}

In the present section, we provide theoretical $M$--$L$ relations for
very massive stars. These relations can be used to estimate masses
$M$, and Eddington factors $\Gamma_{\rm e}$ for observed stars with
known stellar parameters $L$, $T_\star$, and $X_{\rm H}^s$.

A basic problem of such an approach is that the internal structure of
a single observed star is generally not known, i.e., its precise mass
cannot be uniquely predicted. In the present section we thus focus on
the extremes, namely the lowest and highest masses for a star with
given observed parameters. Applying these relations to {\changedA a
  large sample of stars}, we will be able to perform a {\em
  qualitative} investigation of their mass-loss properties, and to
examine patterns that are related to the stellar core, and surface
abundances.

The highest possible mass for a star with given luminosity $L$, and
hydrogen surface abundance $X_{\rm H}^s$ is reached by {\em
  chemically-homogeneous} stars. Under this assumption, the star is
characterized by one, constant hydrogen abundance $X_{\rm H}$, which
equals the surface abundance $X_{\rm H}^s$. The estimated stellar mass
$M_{\rm hom}(L,X_{\rm H}^s)$ is strongly dependent on the {\changedA
  (observable)} surface abundance $X_{\rm H}^s$. Stars that are not
homogeneous have a higher mean molecular weight in the core than at
the surface. They thus have higher $L/M$ ratios, or lower masses, for
given $L$, and $X_{\rm H}^s$.

The lowest possible mass is given by the completely inhomogeneous
case, i.e., by a core hydrogen abundance $X_{\rm H}^c=0$. In this case
the star is in the {\em core He-burning} phase\footnote{Note that we
  only focus on stars in the phase of central H, or He-burning because
  the burning timescales in later phases become extremely short.}.
\citet{lau1:71} have investigated how luminosities, and temperatures
of core He-burning stars vary, depending on the mass ratio between the
H-rich envelope and the He core. According to this work, there exists
a generalized main sequence that depends on the size of the He core.
Stars with very small He cores display hot temperatures, and similar
luminosities as their H-burning counterparts. For larger cores,
temperatures become very cool, and luminosities increase with
increasing core size.  Finally, stars with large He cores show the
same luminosities as pure He-stars, and hot temperatures. Because we
are interested in hot, massive stars with large convective cores in
this work, we can thus use the $M$--$L$ relation for {\em pure}
He-stars for the core He-burning case.  Our minimum mass is thus given
by $M_{\rm Heb}(L) \equiv M_{\rm hom}(L,X_{\rm H}=0)$.  Note that this
mass estimate is completely independent of the (observed) surface
abundance $X_{\rm H}^s$.

The different dependence on $X_{\rm H}^s$ in both cases is of
paramount importance for the present work. Because of the expected
relation between mass and mass loss, we will be able to identify
patterns in the observed properties of very massive stars that help to
distinguish between samples of well-mixed stars, that are
quasi-chemically homogeneous, and stars with a pronounced chemical
profile.

\subsection{The $M$-$L$ relation for homogeneous stars}
\label{sec:Mhom}

\begin{figure}[tp]
  \parbox[b]{0.49\textwidth}{\includegraphics[scale=0.445]{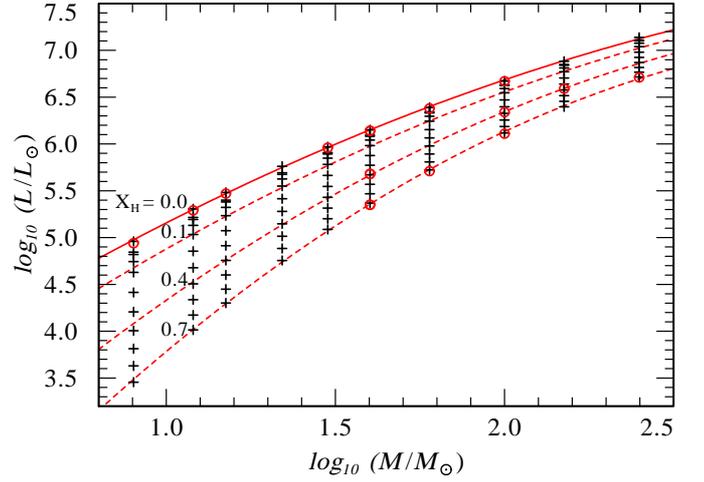}}
  \caption{Homogeneous stellar structure models for the mass range
    8--250\,$M_\odot$ (black symbols). The models are computed for
    hydrogen mass fractions $X_{\rm H} = 0.7$--0 (from bottom to top).
    Dashed red lines indicate the $M$-$L$ relations {\changedC
      according to Eq.\,\ref{eq:MLH_TXT} \& row No.\,1 in
      Table\,\ref{tab:coeff}} for $X_{\rm H} = 0.7$, {\changedA 0.4},
    and 0.1, and the solid red line corresponds to pure He models
    {\changedC according to Eq.\,\ref{eq:MLHE_TXT} \& row No.\,6 in
      Table\,\ref{tab:coeff}}. For comparison, model computations from
    \citet{ish1:99} are indicated by red {\changedA circles}.{
      \changedA The fitting relations are inferred for the mass ranges
      of 12--250\,$M_\odot$ (Eq.\,\ref{eq:MLH_TXT}), and
      8--250\,$M_\odot$ (Eq.\,\ref{eq:MLHE_TXT}).}}
  \label{fig:MLhom}
\end{figure}

\begin{figure}[tp]
  \parbox[b]{0.49\textwidth}{\includegraphics[scale=0.445]{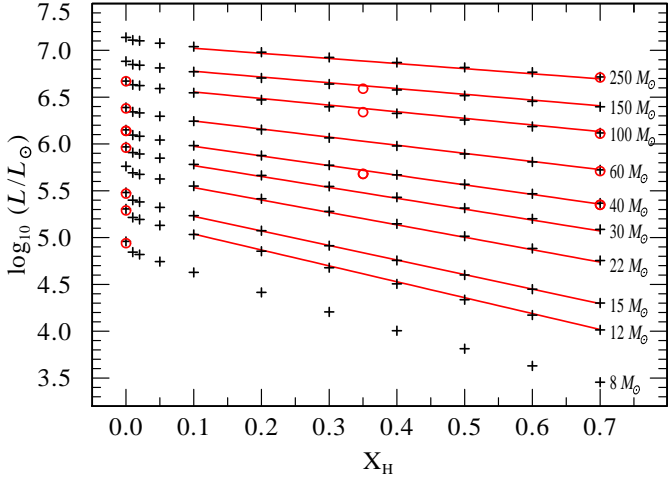}}
  \caption{$M$-$L$ relation for chemically-homogeneous stars,
    dependence on $X_{\rm H}$. The symbols are the same as in
    Fig.\,\ref{fig:MLhom}. Red lines indicate our linear fit to models
    with constant mass but varying hydrogen mass fraction $X_{\rm H}$,
    {\changedA for a mass range of 12--250\,$M_\odot$}, {\changedC
      according to Eq.\,\ref{eq:MLH_TXT} \& row No.\,1 in
      Table\,\ref{tab:coeff}}.}
  \label{fig:MLH}
\end{figure}

In the present section we derive analytical expressions {\changedB of
  the form $L(M,X_{\rm H})$, and $M(L,X_{\rm H})$, for the masses and
  luminosities of chemically-homogeneous stars. As discussed above, we
  expect the luminosity $L$ to depend on the mass $M$, and the
  hydrogen mass fraction $X_{\rm H}$ for core H-burning stars, and
  only on $M$ for core He-burning stars.} To investigate this
dependence we have computed a grid of homogeneous stellar structure
models for the mass range {\changedA $M = 0.3$}--4000\,$M_\odot$, and
hydrogen mass fractions $X_{\rm H} = 0.0$--0.7, at solar metallicity.

The stellar structure models are computed with a simple code that
integrates the stellar structure equations using a shooting method.
The numerical methods are described in the textbook by
\citet{han1:94}.  The code is based on an example program that is
distributed with the book, but is completely re-written, and updated
with OPAL opacities \citep{igl1:96}. To circumvent numerical problems
with the extended envelopes of extremely massive stars
\citep[see][]{ish1:99, pet1:06}, we have adopted an outer boundary
temperature that lies above the temperature of the Fe-opacity peak
($\sim$\,160\,kK). In this way we start our computations just below
the extended envelopes. Because the masses of such envelopes are very
small ($<$\,$10^{-2}\,M_\star$) this approach has no effect on the
obtained luminosities.

The results of our grid computations are displayed in
Figs.\,\ref{fig:MLhom}\,\&\,\ref{fig:MLH}. They are in good agreement
with previous models by \citet{lan1:89}, and \citet{ish1:99} that
comprise a much smaller parameter range. {\changedA We find that a
  polynomial relation that is quadratic in $\log(M)$, and linear in
  $X_{\rm H}$ fits the results satisfactorily} over the parameter
range $M = 12$--250\,$M_\odot$, and $X_{\rm H} = 0.1$--0.7 (see
Figs.\,\ref{fig:MLhom}\,\&\,\ref{fig:MLH}). The resulting relation has
the form
{\changedA
\begin{eqnarray}
\label{eq:MLH_TXT}
  \log(L/L_\odot) & = & [F_1 + F_2\,X_{\rm H}]\\
  \nonumber 
                  & + & [F_3 + F_4\,X_{\rm H}]\,\log(M/M_\odot) \\
  \nonumber
                  & + & [F_5 + F_6\,X_{\rm H}]\,\log(M/M_\odot)^2,
\end{eqnarray}
with the coefficients $F_1$--$F_6$ from row {\changedC No.\,1} in
Table\,\ref{tab:coeff}.}  For $X_{\rm H} < 0.1$, the changes in the
core temperature become so large that the dependence on $X_{\rm H}$
becomes significantly non-linear. We thus restrict our fit to $X_{\rm
  H} > 0.1$, and derive a separate relation for pure He stars ($X_{\rm
  H} =0$), {\changedA over a mass range of $8$--250\,$M_\odot$}. This
relation has the form
{\changedA
  \begin{eqnarray}
    \label{eq:MLHE_TXT}
    \log(L/L_\odot) & = & F_1 + F_2\,\log(M/M_\odot) + F_3\,\log(M/M_\odot)^2,
  \end{eqnarray}
  where the coefficients $F_1$--$F_3$ are again given in {row
    \changedC No.\,6 in} Table\,\ref{tab:coeff}.}  The maximum fitting
error for $\log(L/L_\odot)$ over the given parameter ranges amounts to
0.02.  Inverting these relations we obtain the masses $M_{\rm
  hom}(L,X_{\rm H})$, and $M_{\rm Heb}(L)$ in the form
    \begin{equation}
      \label{eq:LMH_TXT1}
      \log(M_{\rm hom}/M_\odot) = \frac{F_1 + F_2\,X_{\rm H} + F_3 \sqrt{f}}{1 + F_9\,X_{\rm H}}\\
    \end{equation}
    with
    \begin{equation}
      \label{eq:LMH_TXT2}
      f = F_4 + F_5\,X_{\rm H} + F_6\,X_{\rm H}^2
      +(F_7 + F_8\,X_{\rm H})\,\log(L/L_\odot),
    \end{equation}
    and
    \begin{equation}
      \label{eq:LMHE_TXT}
      \log(M_{\rm Heb}/M_\odot) = F_1 + F_2\, \sqrt{F_3 + F_4 \log(L/L_\odot)}.
    \end{equation}
    All coefficients are given in Table\,\ref{tab:coeff},
      together with coefficients for additional relations that cover
      higher, and lower mass ranges. The latter are discussed in
      Appendix \ref{sec:appendix1}, and may be useful for future
      studies.

\section{The most massive stars in the Arches cluster}
\label{sec:arches}

In the present section we confront the concept of $\Gamma$-dependent
mass-loss rates with observations.  To this end we use the large, and
well studied sample of very massive stars in {\changedB the} Arches
cluster, near the Galactic centre. The core of this young massive
cluster contains 13 extremely luminous WNh stars, and a similar amount
of bright early-type O, and Of stars. {\changedA We adopt the stellar
  parameters {\changedB $L$, $X_{\rm H}^s$, $T_\star$, and $\dot{M}$,
    as} obtained by \citet{mar1:08}\footnote{\changedB Note that the
    effective core temperature $T_\star$, given by \citet{mar1:08}, is
    defined in the same way as by \citet{gra1:08}, as the effective
    temperature related to the inner boundary radius $R_\star$ of the
    model atmosphere, i.e., by the relation $L=4\pi R_\star^2\,\sigma
    T_\star^4$, where $R_\star$ is located at large optical depth.
    Due to the small density scale height in these layers, the precise
    value of the reference optical depth has almost no influence on
    the value of $T_\star$, so that the temperatures given by
    \citeauthor{mar1:08} are ideally suited for the use with the
    relation by \citeauthor{gra1:08}.} in a comprehensive study of the
  Arches sample.}

In Sect.\,\ref{sec:Aprop} we start with a description of the
properties of the Arches cluster stars. {\changedC Using a fitting
  technique, we investigate in Sect.\,\ref{sec:fits}, whether the
  observed properties of the Arches stars are in line with a general
  $\Gamma$-dependent mass loss relation.  In Sect.\,\ref{sec:Mwind} we
  determine wind masses from the mass-loss relations by
  \citet{gra1:08}, to test if the theoretical relations cover a
  realistic parameter range.}

\subsection{Properties of the most massive stars in the Arches
  cluster}
\label{sec:Aprop}

\begin{table*} \caption{Stellar parameters of the most massive stars
      in the Arches Cluster.  \label{tab:stars}} \centering
    \begin{tabular}{llllllllllllllll} \hline \hline
      \rule{0cm}{2.2ex}Star & subtype&  $T_\star$ & $L$   & $\log(\dot{M})$                  & $\varv_\infty$           & $X_{\rm H}^s$ & $X_{\rm N}^s$ & $M_{\rm Heb}$ & $M_{\rm hom}$ & $M_{\rm w2}$ & $M_{\rm w1}$ & $\Gamma_{\rm e}^{\rm Heb}$   & $\Gamma_{\rm e}^{\rm hom}$ & $\Gamma_{{\rm e}}^{\rm w1}$ & $\Gamma_{{\rm e}}^{\rm w2}$ \\
      &        &  [kK]      & [$L_\odot$] & [$\frac{M_\odot}{\rm yr}$] & [$\frac{\rm km}{\rm s}$] &             & [\%]        & [$M_\odot$]    & [$M_\odot$]   & [$M_\odot$]    & [$M_\odot$]    & & \\
      \hline \rule{0cm}{2.2ex}%
      F8   &WN8-9   &33.7 &6.10 &$-$4.50 &1000 &0.18 &1.64 &37.1  &55.5  &51.0  &43.3  &0.63  &0.42  &0.46  &0.54  \\
      F5   &WN8-9   &35.8 &5.95 &$-$4.64 &900  &0.22 &1.95 &29.5  &47.3  &36.5  &31.1  &0.58  &0.36  &0.46  &0.54  \\
      F3   &WN8-9   &29.9 &6.10 &$-$4.60 &800  &0.27 &2.79 &37.1  &62.7  &63.2  &52.4  &0.68  &0.40  &0.40  &0.48  \\
      F4   &WN7-8   &37.3 &6.30 &$-$4.35 &1400 &0.36 &2.10 &51.0  &93.6  &75.7  &66.4  &0.83  &0.45  &0.56  &0.64  \\
      F2   &WN8-9   &34.5 &6.00 &$-$4.72 &1400 &0.40 &1.43 &31.8  &63.3  &49.2  &41.6  &0.68  &0.34  &0.44  &0.52  \\
      F7   &WN8-9   &33.7 &6.30 &$-$4.60 &1300 &0.43 &1.86 &51.0  &101.2 &102.3 &86.3  &0.88  &0.44  &0.44  &0.52  \\
      F6   &WN8-9   &34.7 &6.35 &$-$4.62 &1400 &0.54 &1.14 &55.4  &121.8 &119.2 &101.0 &0.97  &0.44  &0.45  &0.53  \\
      F12  &WN7-8   &37.3 &6.20 &$-$4.75 &1500 &0.54 &2.26 &43.4  &97.5  &82.1  &70.0  &0.87  &0.39  &0.46  &0.54  \\
      B1   &WN8-9   &32.2 &5.95 &$-$5.00 &1600 &0.69 &2.41 &29.5  &82.0  &60.4  &49.8  &0.80  &0.29  &0.39  &0.47  \\
      F1   &WN8-9   &33.7 &6.30 &$-$4.70 &1400 &0.69 &1.45 &51.0  &133.0 &119.4 &100.9 &1.03  &0.40  &0.44  &0.52  \\
      F9   &WN8-9   &36.8 &6.35 &$-$4.78 &1800 &0.69 &1.46 &55.4  &143.6 &131.3 &111.3 &1.07  &0.41  &0.45  &0.53  \\
      F14  &WN8-9   &34.5 &6.00 &$-$5.00 &1400 &0.69 &0.49 &31.8  &87.4  &64.8  &54.0  &0.83  &0.30  &0.41  &0.49  \\
      F16  &WN8-9   &32.4 &5.90 &$-$5.11 &1400 &0.69 &1.46 &27.4  &76.9  &56.0  &45.9  &0.76  &0.27  &0.37  &0.46  \\
      \hline \rule{0cm}{2.2ex}%
      F10  &O4-6If  &32.4 &5.95 &$-$5.30 &1600 &0.69 &0.39 &29.5  &82.0  &69.1  &55.3  &0.80  &0.29  &0.34  &0.42  \\
      F15  &O4-6If  &35.8 &6.15 &$-$5.10 &2400 &0.69 &0.49 &40.1  &107.0 &96.9  &79.7  &0.93  &0.35  &0.38  &0.47  \\
      F18  &O4-6I   &37.3 &6.05 &$-$5.35 &2150 &0.69 &0.39 &34.3  &93.3  &82.5  &66.9  &0.86  &0.32  &0.36  &0.44  \\
      F20  &O4-6I   &38.4 &5.90 &$-$5.42 &2850 &0.69 &0.30 &27.4  &76.9  &57.4  &46.7  &0.76  &0.27  &0.36  &0.45  \\
      F21  &O4-6I   &35.8 &5.95 &$-$5.49 &2200 &0.69 &0.39 &29.5  &82.0  &70.2  &56.0  &0.80  &0.29  &0.33  &0.42  \\
      F22  &O4-6I   &35.8 &5.80 &$-$5.70 &1900 &0.69 &0.39 &23.7  &68.0  &52.5  &41.3  &0.70  &0.24  &0.32  &0.40  \\
      F23  &O4-6I   &35.8 &5.80 &$-$5.65 &1900 &0.69 &0.69 &23.7  &68.0  &51.5  &40.7  &0.70  &0.24  &0.32  &0.41  \\
      F26  &O4-6I   &39.8 &5.85 &$-$5.73 &2600 &0.69 &0.40 &25.5  &72.3  &56.6  &45.0  &0.73  &0.26  &0.33  &0.42  \\
      F28  &O4-6I   &39.8 &5.95 &$-$5.70 &2750 &0.69 &0.40 &29.5  &82.0  &71.5  &56.8  &0.80  &0.29  &0.33  &0.41  \\
      F29  &O4-6I   &35.7 &5.75 &$-$5.60 &2900 &0.69 &0.30 &22.1  &64.1  &44.8  &35.6  &0.67  &0.23  &0.33  &0.42  \\
      F32  &O4-6I   &40.8 &5.85 &$-$5.90 &2400 &0.69 &0.29 &25.5  &72.3  &59.3  &46.6  &0.73  &0.26  &0.31  &0.40  \\
      F33  &O4-6I   &39.8 &5.85 &$-$5.73 &2600 &0.69 &0.39 &25.5  &72.3  &56.6  &45.0  &0.73  &0.26  &0.33  &0.42  \\
      F34  &O4-6I   &38.1 &5.75 &$-$5.77 &1750 &0.69 &0.40 &22.1  &64.1  &46.0  &36.3  &0.67  &0.23  &0.32  &0.40  \\
      F35  &O4-6I   &33.8 &5.70 &$-$5.76 &2150 &0.69 &0.20 &20.6  &60.5  &43.1  &33.6  &0.64  &0.22  &0.31  &0.39  \\
      F40  &O4-6I   &39.8 &5.75 &$-$5.75 &2450 &0.69 &0.40 &22.1  &64.1  &44.5  &35.5  &0.67  &0.23  &0.33  &0.42  \\
      \hline \end{tabular}   \tablefoot{ Designations,
      subtypes, stellar temperatures ($T_\star$), luminosities ($L$),
      mass-loss rates ($\dot{M}$), terminal wind velocities
      ($\varv_\infty$), hydrogen, and nitrogen surface mass fractions
      ($X_{\rm H}^s$, $X_{\rm N}^s$), from \citet{mar1:08}, and mass
      estimates from the present work: He-burning masses ($M_{\rm Heb}$),
      and homogeneous masses ($M_{\rm hom}$) according to {\changedC
        Eqs.\,\ref{eq:LMH_TXT1}--\ref{eq:LMHE_TXT}, with coefficients
        from rows 11, \& 16 in Table\,\ref{tab:coeff}}, wind masses
      according to \citet{gra1:08} for $Z=2\,Z_\odot$ ($M_{\rm w2}$), and
      $Z=1\,Z_\odot$ ($M_{\rm w1}$). The corresponding Eddington factors
      $\Gamma_{\rm e}$ are given in the last four columns.  }
\end{table*} 

The stellar parameters of the Arches stars, as derived by
\citet{mar1:08}, are {\changedA compiled} in Table\,\ref{tab:stars}.
Spectral types reach from early O, and Of supergiants to late WNh
subtypes. Stellar temperatures lie in the range $T_\star =
30$--40\,kK, and luminosities in the range $\log(L_\star/L_\odot) =
5.75$--6.35, with the WNh stars showing systematically higher
luminosities, and lower temperatures. Moreover, the large part of the
WNh stars is H-deficient and N-enriched, with respect to solar values.
Note, however, that some WNh stars display a solar hydrogen abundance
at their surface.

Based on the derived mass-loss rates, \citet{mar1:08} identify two
distinct wind momentum -- luminosity relations for the O/Of, and WNh
stars.  The WNh stars display systematically higher mass-loss rates
than the O stars.  This raises the question in which way the WNh stars
differ from the O/Of stars. On the one hand, the different mass-loss
properties could indicate a different evolutionary stage, e.g., the
WNh stars could be in the phase of core He-burning. On the other hand,
parameters like surface abundances, or effective temperatures could be
responsible for the observed dichotomy.

\subsection{\changedC{$\Gamma$-dependent mass-loss rates}}
\label{sec:fits}

\begin{figure}[tp]
\parbox[b]{0.49\textwidth}{\includegraphics[scale=0.435]{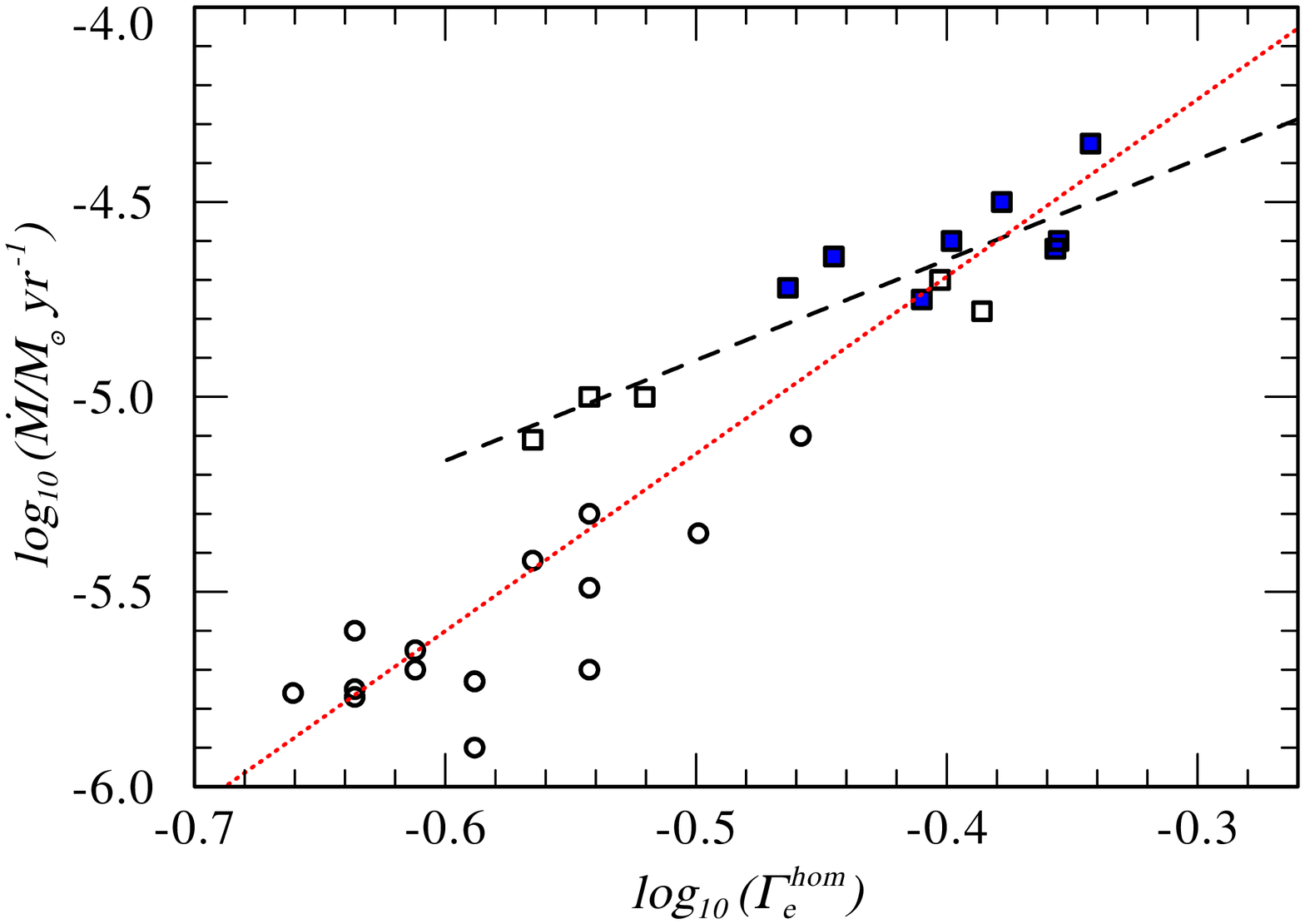}}  
\caption{Under the assumption of chemical homogeneity, the Arches
  cluster stars display a pronounced $\Gamma$-dependence.  {\changedA
    Squares indicate stars with spectral subtypes WNh (filled blue:
    H-deficient; unfilled: normal H abundance), and circles O/Of.}
  The black {\changedA dashed} line indicates the fitted mass-loss
  relation (Eq.\,\ref{eq:massloss2}) for WNh stars (parameters from
  Table\,\ref{tab:fit} row 2). The red {\changedA dotted} line indicates
  the {\changedA corresponding} fit for the complete sample (parameters
  from Table\,\ref{tab:fit2} row 2).  }
  \label{fig:gamma1}
\end{figure}

\begin{figure}[tp]
\parbox[b]{0.49\textwidth}{\includegraphics[scale=0.435]{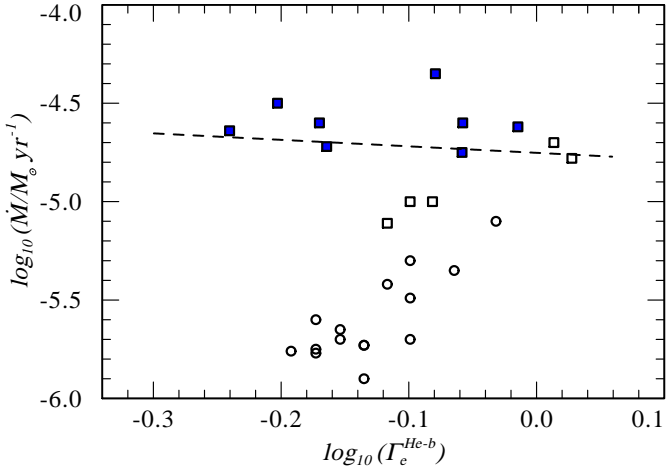}}  
\caption{Under the assumption of central He-burning, the Arches
  cluster stars display no general $\Gamma$-dependence. Symbols are
  the same as in Fig.\,\ref{fig:gamma1}, the black {\changedA dashed}
  line indicates a fitted relation (Eq.\,\ref{eq:massloss2}) for WNh
  stars (parameters from Table\,\ref{tab:fit} row 3).}
  \label{fig:gamma2}
\end{figure}

\begin{table}[tp] \caption{Results of various $\chi^2$ fits of
    Eq.\,\ref{eq:massloss2} to the observed mass-loss rates of the
    13 WNh stars in the Arches sample.\label{tab:fit} } \centering
  \begin{tabular}{l|lllllllllllllll} \hline \hline
    \rule{0cm}{2.2ex}$\chi^2$ & $\dot{M}_0$ &  $f_\Gamma^{\rm hom}$ &  $f_\Gamma^{\rm Heb}$ & $f_L$ & $f_T$ \\
    \hline
    \rule{0cm}{2.2ex}0.544 & $-$4.721 & $-$    & $-$    & $-$    & $-$ \\
    \hline
    \rule{0cm}{2.2ex}0.108 & $-$3.617 & 2.578 & $-$    & $-$    & $-$ \\
    0.536 & $-$4.752 & $-$    & $-$0.330 & $-$    & $-$ \\
    0.368 & $-$4.820 & $-$    & $-$    & 0.720 & $-$ \\
    0.531 & $-$4.751 & $-$    & $-$    & $-$    & 1.204 \\
    \hline
    \rule{0cm}{2.2ex}0.354 & $-$4.800 & $-$    & $-$    & 0.843 & $-$1.438 \\
    0.039 & $-$2.802 & 4.208 & $-$    & $-$0.845 & $-$ \\
    0.098 & $-$3.537 & 2.703 & $-$    & $-$    & $-$1.056 \\
    0.037 & $-$5.244 & $-$    & $-$2.937 & 1.749 & $-$ \\
    0.496 & $-$4.859 & $-$    & $-$0.787 & $-$    & 2.467 \\
    \hline
    \rule{0cm}{2.2ex}0.037 & $-$2.768 & 4.293 & $-$    & $-$0.920 & 0.495 \\
    0.035 & $-$5.262 & $-$    & $-$3.008 & 1.724 & 0.577 \\
        \hline
      \end{tabular}
      \tablefoot{Parameters indicated by ($-$) have been set to zero.}
\end{table}

{\changedC In this section we investigate whether the observed
  properties of the Arches cluster stars are in line with a
  $\Gamma$-dependent mass-loss relation. For this purpose we adopt a
  general mass loss relation in the form of a power law} with four
free parameters
\begin{equation}
 \label{eq:massloss2}
  \begin{array}{l}
    \log(\dot{M}/{\changedA \msunpyr}) =   \dot{M}_0 + f_\Gamma \cdot \log(\Gamma_{\rm e}) \\
    \rule{0cm}{2.4ex}\hspace{0.75cm} 
    + f_L \cdot \left(\log(L/L_\odot)        - 6.0\right) 
    + f_T \cdot \left(\log(T_\star/{\rm K})  - 4.5\right).
  \end{array}
\end{equation} 

{\changedC To estimate $\Gamma_{\rm e}$, we use Eq.\,\ref{eq:gamma},
  with the relations for $M_{\rm hom}(L,X_{\rm H})$, and $M_{\rm
    Heb}(L)$ from Sect.\,\ref{sec:Mhom}. This way, it is possible to
  constrain the free parameters $\dot{M}_0$, $f_\Gamma$, $f_L$, and
  $f_T$ by a $\chi^2$-fit, based on the observed values of $\dot{M}$,
  $L$, $T_\star$, and $X_{\rm H}^s$.

  The hydrogen surface abundance $X_{\rm H}^s$ plays a crucial role in
  this fitting process, as it enters the estimate of $\Gamma_{\rm e}$
  via $\chi_{\rm e}$ (Eq.\,\ref{eq:gamma}), and via $M_{\rm
    hom}(L,X_{\rm H})$. The wide spread in surface abundances for the
  WN stars in our sample is important for the verification of an
  actual $\Gamma$-dependence. Without variations in $X_{\rm H}^s$, a
  $\Gamma$-dependence could not be easily distinguished from an
  $L$-dependence, if there exists an underlying $M(L)$ relation that
  does not depend on other parameters.
  In the following we will see that this may indeed happen for the O
  stars within our sample, which all have the same $X_{\rm H}^s$.}

Using Eq.\,\ref{eq:massloss2}, we perform $\chi^2$-fits to the
observed mass-loss rates of the sample stars, {\changedC based on the
  observed values of $L$, $T_\star$, and $X_{\rm H}^s$. Dependent on
  whether $\Gamma_{\rm e}$ is estimated on the basis of $M_{\rm
    Heb}(L)$, or $M_{\rm hom}(L,X_{\rm H})$, we designate the obtained
  values of $f_\Gamma$ in Eq.\,\ref{eq:massloss2} as $f_\Gamma^{\rm
    hom}$, or $f_\Gamma^{\rm Heb}$.} To extract the physically
relevant parameters we perform a series of tests where the fitting
parameters $f_\Gamma$, $f_L$, and $f_T$ are partly set to zero.

We start our investigation using only the 13 WNh stars in the Arches
sample.  The results of our $\chi^2$-fits are {\changedA compiled} in
Table\,\ref{tab:fit}.  We start with the most simple case of a constant
mass-loss rate, i.e., with $\dot{M}_0$ as the only free parameter. The
obtained $\chi^2=0.544$ obviously marks the lower end of the
achievable fit quality. In the next four rows of Table\,\ref{tab:fit}
we allow for one more free parameter apart from $\dot{M}_0$. Among
those four cases the best result is obtained for a $\Gamma$-dependent
mass-loss relation under the assumption of homogeneity. Note the big
difference in $\chi^2$ between the homogeneous assumption, and the
assumption of core He-burning.  {\changedA This is illustrated in
  Figs.\,\ref{fig:gamma1} \& \ref{fig:gamma2}, where the sample stars
  indeed} form a well pronounced mass-loss relation under the
assumption of homogeneity (Fig.\,\ref{fig:gamma1}), but not under the
assumption of He-burning (Fig.\,\ref{fig:gamma2}). {\changed At this
  point we note that the two plots represent two extremes, and that
  the real $\dot{M}$--$\Gamma$ relation may lie in between the two
  cases.}

The qualitative difference between {\changed the} two plots originates
from the fact that the adopted $M$--$L$ relation in
Fig.\,\ref{fig:gamma1} depends on $X_{\rm H}^s$, while the relation in
Fig.\,\ref{fig:gamma2} does not. In both plots the O stars form a
separate sequence that merges into the WR sequence. As explained
above, this does not necessarily indicate a true $\Gamma$-dependence
for the O star sample, as these stars show no variations in $X_{\rm
  H}^s$.  The WNh stars, on the other hand, display a substantial
spread in $X_{\rm H}^s$.  Notably, in Fig.\,\ref{fig:gamma2}, the
H-deficient stars show a significant shift towards lower $\Gamma_{\rm
  e}$, {\changed i.e., a 'hook' in the $\dot{M}$--$\Gamma$ relation.}
This can be explained by the dependence of $\Gamma_{\rm e}$ on $X_{\rm
  H}^s$ in Eq.\,\ref{eq:gamma}, with $\Gamma_{\rm e} \propto 1+X_{\rm
  H}^s$.  The fact that this effect is precisely compensated in
Fig.\,\ref{fig:gamma1}, suggests that the internal structure of the
WNh stars in our sample is {\changed close to homogeneity.  More
  precisely, it shows that there exists a relation between core, and
  surface hydrogen abundance.  A similar effect may thus occur for
  inhomogeneous stars that have a similar degree of inhomogeneity, so
  that the differential behaviour of the homogeneous case is
  preserved.  However, also in this case, the largest part of the
  sample stars need to be in phase of core H-burning to display the
  observed behaviour. Only few individual objects, with the highest He
  surface enrichment, may already have reached the He-burning phase.}

\begin{figure}[tp]
\parbox[b]{0.49\textwidth}{\includegraphics[scale=0.435]{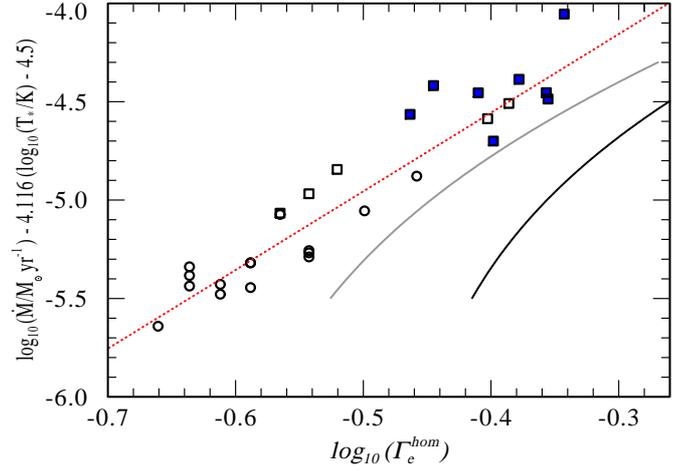}}  
\caption{Fitted mass-loss relation for the complete Arches sample,
  under the assumption of chemical homogeneity {\changedA
    (Eq.\,\ref{eq:mdotfit}, red dotted line)}. Symbols are the same as
  in Fig.\,\ref{fig:gamma1}.  The observed mass-loss rates are scaled
  to a temperature of $\log(T_\star/{\rm K})=4.5$, based on the
  temperature dependence {\changedA in Eq.\,\ref{eq:mdotfit},} with
  $\dot{M}\propto T^{-4.116}$.  For comparison, we plot the relation
  by \citet{gra1:08} for the same temperature, $\log(L)=6.3$, $X_{\rm
    H}=0.7$, and $Z=Z_\odot$ (black solid), as well as $Z=2Z_\odot$
  (grey solid).}
  \label{fig:gamma3}
\end{figure}

\begin{table}[tp]
  \caption{Results of various $\chi^2$
    fits of Eq.\,\ref{eq:massloss2} to the observed mass-loss
    rates of the complete Arches sample.\label{tab:fit2}}
 \centering
 \begin{tabular}{l|lllllllllllllll} \hline \hline
   \rule{0cm}{2.2ex}$\chi^2$ & $\dot{M}_0$ &  $f_\Gamma^{\rm hom}$ & $f_L$ & $f_T$ \\
   \hline
   \rule{0cm}{2.2ex}6.557 & $-$5.190 & $-$     & $-$    & $-$ \\
   \hline
   \rule{0cm}{2.2ex}0.942 & $-$2.871 & 4.550  & $-$    & $-$ \\
   2.037 & $-$5.179 & $-$     & 2.068 & $-$ \\
   4.160 & $-$4.880 & $-$     & $-$    & $-$8.458 \\
   \hline
   \rule{0cm}{2.2ex}0.783 & $-$1.694 &  6.872  & $-$1.239 & $-$ \\
   0.457 & $-$2.957 & 3.997  & $-$    & $-$4.116 \\
   \hline
   \rule{0cm}{2.2ex}0.429 & $-$2.420 & 5.091  & $-$0.557 & $-$3.751 \\
   \hline
 \end{tabular}
 \tablefoot{Parameters indicated by ($-$) have been set to zero.}
\end{table} 

In the remainder of Table\,\ref{tab:fit} we allow for three or four
free parameters.  For these cases similar fitting results are obtained
for the He-burning, and the homogeneous case.  However, the obtained
values for $f_\Gamma^{\rm Heb}$ are all negative. Such a relation
would imply that mass loss ceases close to the Eddington limit, which
seems to be rather unphysical. An inspection of Fig.\,\ref{fig:gamma2}
shows that the assumption of He-burning would indeed imply a slightly
negative slope of the mass-loss relation for WNh stars. The fact that
for this case a similar fit quality can be achieved as for the
homogeneous case is presumably a sign that the number of fitting
parameters becomes too large for the small sample.

In Fig.\,\ref{fig:gamma1} it is notable that our fitted
$\dot{M}(\Gamma_{\rm e})$ relation for WNh stars (black dashed line in
Fig.\,\ref{fig:gamma1}) seems to be slightly detached from the rest of
the sample, similar to the two different mass-loss relations for WN
and O/Of stars identified by \citet{mar1:08}.  Fig.\,\ref{fig:gamma1},
however, suggests that there might exist a smooth transition between
the WNh, and O/Of stars {\changedA \citep[cf.][]{vin1:11}}. To
investigate this possibility we have performed the same fitting
procedure as previously, for the complete Arches sample.  Because the
O stars are unlikely to be in the phase of core He-burning, we have
however only focused on the homogeneous case.

The results are combined in Table\,\ref{tab:fit2}.  Again, we start
with the {\changedC simplest} case of a constant mass-loss rate
$\dot{M}_0$ with $\chi^2=6.557$.  The next three rows allow for one
more free parameter. Again, a $\Gamma$-dependent mass-loss relation is
strongly favored with $\chi^2=0.942$. Interestingly, the following
fits with more free parameters strongly favor a temperature dependence
with $\dot{M}\propto T^{-4}$, which is very similar to the dependence
predicted by \citet{gra1:08}.  The two fits with the lowest $\chi^2$
in Table\,\ref{tab:fit2} do generally support the key features of
{\changedA their} mass-loss predictions, which are 1) a strong
$\Gamma$-dependence, 2) a weak luminosity dependence, and 3) a strong
increase in mass loss for decreasing $T_\star$.

From row 6 in Table\,\ref{tab:fit2} we obtain the mass-loss relation
\begin{equation}
 \label{eq:mdotfit}
  \begin{array}{l}
    \log(\dot{M}/{\changedA \msunpyr}) =  \\
    \rule{0cm}{2.4ex}\hspace{0.3cm} 
    - 2.957
    + 3.997 \cdot \log(\Gamma_{\rm e})
    - 4.116 \cdot \left(\log(T_\star/{\rm K})-4.5\right).
  \end{array}
\end{equation} 
In Fig.\,\ref{fig:gamma3} we compare this relation with observed
mass-loss rates, compensating for the temperature dependence.
Notably, the previous distinction between the O/Of, and the WNh sample
from Fig.\,\ref{fig:gamma1} does not exist anymore in this plot.  The
temperature difference between these two samples can thus account for
the differences in their mass-loss rates. {\changedC We note however
  that the dependence on $T_\star$ is uncertain, as it only follows
  from the combined O + WR sample. From the WR sample alone there is
  no evidence for such a dependence, presumably due to the small
  spread in $T_\star$. At this time it is not clear at which point the
  transition to a $\Gamma$-dependent mass loss relation occurs, and to
  which extent such a relation is applicable to (parts of) the O\,star
  sample.}

\begin{figure}[tp]
\parbox[b]{0.49\textwidth}{\includegraphics[scale=0.445]{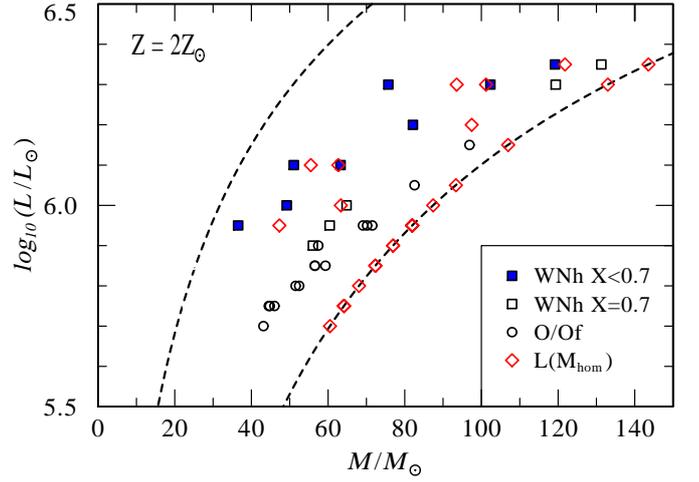}}
\caption{ Mass--luminosity relations $L(M)$ for the most massive stars
  in the Arches cluster, {\changedC for an adopted metallicity of
    $Z=2Z_\odot$}.  Plotted are {\em observed} luminosities, vs.\
  different mass estimates.  Red diamonds indicate $L(M_{\rm hom})$,
  with homogeneous masses $M_{\rm hom}(L,X_{\rm H})$ computed from
  {\changedC Eqs.\,\ref{eq:LMH_TXT1}, \& \ref{eq:LMH_TXT2}, with
    coefficients from row 11 in Table\,\ref{tab:coeff}}.  Black/blue
  symbols indicate $L(M_{\rm wind})$, with wind masses obtained from
  the mass-loss prescription by \citet{gra1:08}
  (Eqs.\,\ref{eq:massloss1}--\ref{eq:fit3}) for $Z=2\,Z_\odot$.
  {\changedA For the latter, we distinguish between WNh subtypes
    (squares), and O/Of subtypes (circles). Among these, blue filled
    symbols indicate H-deficient (WNh) stars.} The dashed lines
  indicate the He-MS (left), and the ZAMS (right) according to our
  relations in Sect.\,\ref{sec:Mhom}.  }
\label{fig:MLarches}
\end{figure}

\begin{figure}[tp]
\parbox[b]{0.49\textwidth}{\includegraphics[scale=0.445]{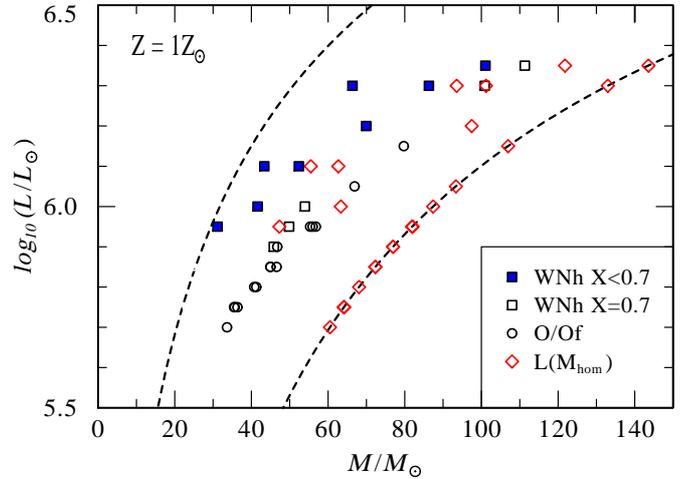}}
\caption{{\changedC Mass--luminosity relations $L(M)$ for the most
    massive stars in the Arches cluster, for an adopted metallicity of
    $Z=1Z_\odot$. Symbols are the same as in Fig.\,\ref{fig:MLarches}.}}
\label{fig:MLarches1Z}
\end{figure}

\subsection{Wind masses for the Arches cluster stars}
\label{sec:Mwind}

{\changedC In the previous section we have shown that the observed
  properties of the Arches cluster stars are in line with a
  $\Gamma$-dependent mass-loss relation. The comparison in
  Fig.\,\ref{fig:gamma3} shows that, under the assumption of chemical
  homogeneity, the resulting mass-loss relation is in good overall
  agreement with the relation by \citet[][]{gra1:08} for a high
  metallicity of $Z\sim 2\,Z_\odot$. In the present section we use
  wind masses that follow from the relation by
  \citeauthor[][]{gra1:08}, to investigate if the Arches stars conform
  with this mass-loss relation, and if the obtained wind masses cover
  a plausible range.}

We derive wind masses by computing the Eddington factor $\Gamma_{\rm
  e}$ that is needed to explain the observed mass-loss rates
$\dot{M}$, for given stellar parameters $L$, $T_\star$, $X_{\rm H}$,
and $Z$, from Eqs.\,\ref{eq:massloss1}--\ref{eq:fit3}. The wind
masses follow from $\Gamma_{\rm e}$, according to
Eq.\,\ref{eq:gamma}.  In Table\,\ref{tab:stars} we list the resulting
masses, and Eddington factors.  The different columns in
Table\,\ref{tab:stars} denote values obtained for an adopted
metallicity of $2Z_\odot$ ($M_{\rm w2}$), and $1Z_\odot$ ($M_{\rm
  w1}$).  The results are compared to homogeneous masses, and
He-burning masses from Sect.\,\ref{sec:Mhom}.

We note that the mass-loss relation by \citet{gra1:08} is obtained for
stellar winds with an optical depth of the order of one, or higher,
and that the O stars with lower luminosities in our sample may just
lie outside the applicable range of this relation.  Nevertheless we
include them here, also to check the validity of the mass-loss
relation.

In Fig.\,\ref{fig:MLarches} we compare the resulting $M$--$L$ relation
{\changedC for $Z=2\,Z_\odot$}, with the expected relation for
homogeneous stars.  We find a notable {\changedB overall} agreement
between both $M$--$L$ relations.  Morphologically, both relations show
two branches in the $M$--$L$ diagram.  The first branch, {\changedB
  with higher luminosities for given wind mass, consists of
  H-deficient stars with subtypes WN7-9h (blue filled symbols).  The
  second branch, with lower luminosities, consists of stars with
  normal hydrogen surface composition (empty black symbols). This
  group covers spectral subtypes O4-6I, O4-6If+, and WN8-9h. Our
  theoretically predicted masses are indicated by red diamonds. Also
  they show two branches, one on the ZAMS (at lower luminosities), and
  one in the region of our previous high-luminosity branch.}

For the theoretically {\changedA predicted} masses $M_{\rm hom}(L,
X_{\rm H}^s)$ (red {\changedA diamonds}) the interpretation is very
simple. By definition, homogeneous stars with solar H-abundance
populate the ZAMS, while H-deficient stars have a higher mean
molecular weight, and thus display higher luminosities.

For the wind masses $M_{\rm w2}$ (black/blue symbols), the difference
{\changedA between} both branches mainly originates from the higher
mass-loss rates of the WNh stars, and the temperature difference
between the WNh, and the O stars.  For the H-deficient {\changedA
  branch, which consists only of WNh subtypes,} the derived wind
masses are in {\changedB very good overall} agreement with {\changedA
  the} homogeneous masses.  For the H-rich {\changedA branch}, the
derived wind masses {\changedC do not fit the homogeneous masses, but
  are} systematically lower, as expected for stars that have just
evolved away from the ZAMS. Again, we have to keep in mind that the
stars with the lowest luminosities in this group, may lie beyond the
applicable regime of {\changedC the theoretical} mass-loss relation.

{\changedC For the H-deficient WNh stars, we only find a good
  quantitative agreement between empirically determined wind masses,
  and theoretically predicted homogeneous masses, if we assume a high
  metallicity of $2\,Z_\odot$.}  {\changedC A similar} metallicity is
indeed measured indirectly for the WNh stars in the Arches sample,
based on their nitrogen abundances.  However, more recent
investigations of stars in the GC, favor a solar value (cf.\
Sect.\,\ref{sec:uncert}).  Using the mass-loss relation by
\citet{gra1:08} with solar $Z$, the derived masses are lower, but the
morphological similarities remain {\changedC (cf.\
  Fig.\,\ref{fig:MLarches1Z}). In this case the wind masses of both
  branches in the diagram imply an inhomogeneous stellar structure.}

We conclude that, using the mass-loss prescription by \citet{gra1:08},
we find two branches in the $M$-$L$ diagram. The more luminous branch
is populated by H-deficient WNh stars. {\changedC Based on an adopted
  metallicity of $2\,Z_\odot$, the derived wind masses for these stars
  would be in agreement with chemically homogeneous stars.}
Interestingly, these stars also show the strongest nitrogen enrichment
(cf.\ Table\,\ref{tab:stars}), which points to a previous evolution
with episodes of strong mixing, {\changedB and/or strong mass loss,
  that could indeed lead to a quasi-homogeneous structure of these
  objects (cf.\,Sect.\,\ref{sec:evolution}).}  The H-rich stars, on
the other hand, populate a branch with lower luminosities, in
agreement with stars that have undergone a normal evolution off the
ZAMS. 

{\changedC At this point we want to note that our preference for a high
  $Z$ could mean that the mass-loss relation by \citet{gra1:08} still
  underestimates the true mass loss rates. In this case the high $Z$
  would compensate for lacking elements in the models. Another
  possibility would be, that the sample stars are not homogeneous, and
  thus $\Gamma_{\rm e}$ is underestimated by our approach.  However,
  in any case the wind masses cover a plausible parameter range.
  Notably, almost all of the sample stars lie below the He
  main-sequence in the $M$--$L$ diagram.  In agreement with our
  findings from Sect.\,\ref{sec:fits}, the largest part of the Arches
  cluster stars is thus most likely still in the phase of core hydrogen
  burning.}

\section{Discussion} 
\label{sec:discussion}

In the present work we have investigated the mass-loss properties of
the most massive stars in the Arches cluster.  We have used a
{\changedA semi-empirical} method that is based on observations, or
more precisely, on the results of spectral analyses, and on
theoretically predicted $M$--$L$ relations.  As a result we found
evidence for the existence of a largely $\Gamma$-dependent mass-loss
relation, and obtained information about the evolutionary status of
the Arches cluster stars.

In the following we discuss our results. We start with the question
how the results are affected by potential uncertainties in the
observed stellar parameters (Sect.\,\ref{sec:uncert}). In
Sect.\,\ref{sec:disc_mdot} we discuss the implications for the mass
loss properties of massive stars, and in Sect.\,\ref{sec:mwind} we
reconsider the potential to obtain precise wind masses. Finally, in
Sect.\,\ref{sec:evolution}, we discuss the implications on the
evolutionary status of the Arches cluster stars.

\subsection{Observational uncertainties}
\label{sec:uncert}

\begin{figure}[tp]
\parbox[b]{0.49\textwidth}{\includegraphics[scale=0.445]{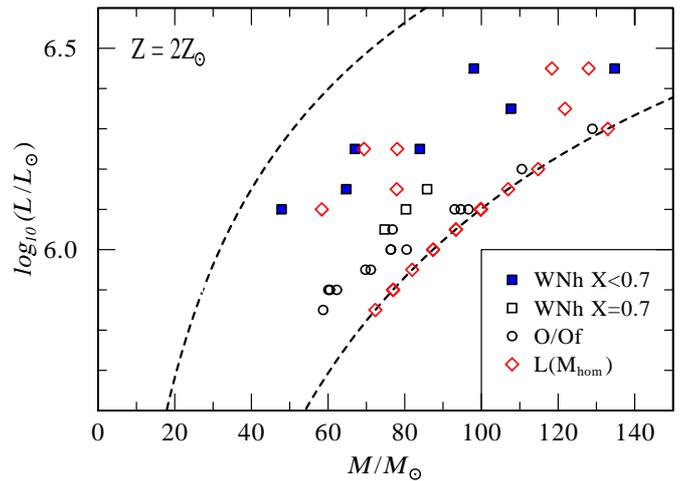}}
\caption{{\changedC Mass--luminosity relations $L(M)$ for the most
    massive stars in the Arches cluster, for an adopted metallicity of
    $Z=2Z_\odot$. To investigate the influence of systematic
    uncertainties on our results, the observed luminosities have been
    artificially increased by 0.15\,dex. To derive wind masses, the
    mass-loss rates have been increased according to $\dot{M}\propto
    L^{3/4}$, as expected for the results from recombination line
    analyses \citep[cf.,][]{ham1:98}.  The symbols are the same as in
    Fig.\,\ref{fig:MLarches}.}}
\label{fig:MLarchesL}
\end{figure}

Here we discuss to which extent our results are affected by
uncertainties in the stellar parameters, that we use as input for our
method. We give an overview of potential error sources, discuss their
quantitative importance, and their influence on the qualitative
outcome of our work. Generally it can be said that, due to the
steepness of the proposed $\dot{M}$--$\Gamma$ relation, our results
are relatively insensitive to such uncertainties.  In addition, most
error sources are of systematic nature, and hardly affect the
qualitative outcome of our work.

\citet{mar1:08} quote uncertainties of $\pm$3\,kK (2\,kK) on $T_{\rm
  eff}$ for O stars (WNLh), $\pm$0.2 on $\log(L)$, 0.2 (0.1) on
$\log(\dot{M})$, and $\pm$50\% ($\pm$30\%) on abundances. If these
(rather conservative) error estimates would be of purely statistical
nature, they would be devastating for our results. The fact that we
actually find well-defined relations for the sample stars, indicates
that potential errors are chiefly of systematic nature. The
possibility that there are differences in the systematics between O
stars, and WNh stars, could however influence our results,
particularly regarding the temperature dependence in
Sect.\,\ref{sec:fits}.

Also \citet{cro1:10} recently pointed out that \citet{mar1:08} may
have systematically under-estimated the stellar luminosities of the
Arches cluster stars, by using a relatively {\changedB small} distance
estimate \citep[$m-M=14.4\pm 0.1$,][]{eis1:05}, and a low foreground
extinction \citep[$A_{\rm K}=2.8\pm 0.1$][]{sto1:02} for the GC.
{\changedA Using} $m-M=14.5\pm 0.1$ \citep{rei1:93}, and $A_{\rm
  K}=3.1\pm 0.1$ \citep{kim1:06} \citeauthor{cro1:10} find slightly
higher luminosities for the two brightest stars in the sample
($\log(L/L_\odot) = 6.5$ instead of 6.35).

Although a systematic increase of $\log(L)$ by 0.15\,dex would be
significant, {\changedC the $\Gamma_{\rm e}$, as derived from the
  mass-loss relations in Sect.\,\ref{sec:pred}}, would hardly be
affected.  The reason is that {\changedC spectroscopic} mass-loss rates
derived from recombination line analyses, as in \citet{mar1:08}, scale
with $\dot{M}\propto L^{0.75}$.  As the luminosity scaling in our
mass-loss relations is very similar (Eqs.\,\ref{eq:vink1},
\ref{eq:vink2}, \& Eq.\,\ref{eq:massloss1}), the derived $\Gamma_{\rm
  e}$ would nearly stay the same. {\changedC This would however still
  result in uncertainties in the derived wind masses, basically with
  $M_{\rm wind} \propto L (1+X_{\rm H}^s)$ (cf.\,Eq.\,\ref{eq:gamma}).
  The effect of a systematic luminosity increase by 0.15\,dex is
  illustrated in Fig.\,\ref{fig:MLarchesL}.}

{\changedC For the mass-loss relations in Sect.\,\ref{sec:fits}, a
  luminosity increase mainly results in a shift of the inferred
  mass-loss rates with $\dot{M}\propto L^{0.75}$. In addition, the
  change of the slope of the $M$--$L$ relation changes the slope of
  the mass-loss relations, with respect to $\Gamma_{\rm e}$.  E.g.,
  based on relation No.\,1 in Table\,\ref{tab:coeff}, an increases of
  0.15 in $\log(L)$ for a star with $\log(L/L_\odot)=6.3$, and $X=0.7$
  (i.e., $\log(\Gamma_{\rm e})=-0.386$) causes an increase of
  $\Gamma_{\rm e}$ by 0.048\,dex.  For a star with
  $\log(L/L_\odot)=5.7$ ($\log(\Gamma_{\rm e})=-0.647$), the same
  increase causes $\Gamma_{\rm e}$ to increase by 0.073\,dex.
  Consequently, the exponents in our empirical mass-loss relations
  increase by a factor of $1+(0.073-0.048)/(0.647-0.386)=1.1$.}

Additional uncertainties in the mass-loss rates are due to wind
clumping (with $\dot{M}\propto \sqrt{f_{\rm cl}}$), and the unknown
wind velocity field. Again, these would introduce a shift in
$\dot{M}$, but only moderately affect the derived $\Gamma_{\rm e}$,
because of the steepness of the proposed $\Gamma$-dependence.  Based
on the aforementioned uncertainties, we thus expect systematic effects
on our quantitative results, particularly concerning mass-loss rates,
and wind masses, but only moderate relative changes, i.e., our results
stay qualitatively the same.

Apart from that, the metallicity $Z$ is of major importance.
{\changedA In contrast to the {\changedB above}, $Z$ does affect the
  $\Gamma$-dependence in our theoretical mass-loss relations, but not
  the observed properties of the sample stars. As discussed in
  Sect.\,\ref{sec:Mwind}, it thus affects the derived wind masses.}
In the relation by \citet[][Eq.\,\ref{eq:massloss1}]{gra1:08}, the
metallicity determines the value $\Gamma_0(Z)$ (Eq.\,\ref{eq:fit2}),
which roughly marks the value of $\Gamma_{\rm e}$ for which the mass
loss starts to increase due to the proximity to the Eddington limit.
Moreover, the exponent $\beta(Z)$ {\changedA defines the steepness of
  the mass-loss relation.  Also \citet{vin1:05} find} a mass-loss
dependence with $\dot{M}\propto Z^{0.68}$ for a typical Wolf-Rayet
model. The reason {\changedA for the strong $Z$-dependence} lies in
the dominant role of Fe line opacities in the radiative acceleration
of hot star winds. The $Z$ dependence in the wind models thus chiefly
reflects a dependence on the Fe abundance.

As mentioned before, the metallicity of the Arches cluster is still
uncertain. It has been determined indirectly, on the basis of observed
nitrogen surface abundances of WNh stars \citep{naj1:04, mar1:08}.
These stars are strongly nitrogen enriched, obviously by material
which has been processed in the CNO cycle in the stellar core.  This
material is exposed at the stellar surface, presumably by mixing
and/or mass loss. The observed N abundance thus provides a lower limit
for the initial abundance of C+N+O.  For the WNh stars in the Arches
cluster \citet{naj1:04} find $X_{\rm N} \approx 1.6$\%, and
\citet{mar1:08} find an average of $X_{\rm N} \approx 1.7$\%, with
values reaching up to $X_{\rm N}=2.79$\%. Both authors conclude that
the initial metallicity of the Arches cluster is only slightly
super-solar.  However, taking into account the recent downward
revision of the solar oxygen abundance \citep[according to][it follows
that $X_{C+N+O, \odot} \approx 0.8$\%]{asp1:05,asp1:09,pei1:09}, a
metallicity of $2\,Z_\odot$ seems more likely.  {\changedC As the
  solar Fe abundance has hardly been affected by this revision, a
  simple scaling would imply that $X_{\rm Fe}/X_{{\rm Fe},\odot} =
  2$.}

We note however that recent spectral analyses of LBVs \citep{naj1:09},
and red supergiants \citep{dav1:09}, {\changedC as well as earlier
  studies of cool stars in the GC region
  \citep{car1:00,ram1:00,cun1:07} indicate a solar Fe abundance, and
  an increased abundance of alpha elements.}  The resulting
uncertainty in the derived wind masses is of the order of 20\% (see
Table\,\ref{tab:stars}). Generally, radiatively driven winds become
stronger for higher $Z$, i.e., the derived wind masses become
systematically larger if a higher $Z$ is adopted.  {\changedA As the
  derived mass ratios are hardly affected, we do not expect a
  qualitative impact on our results from Sect.\,\ref{sec:Mwind}.}
{\changedC The preference for a high Fe abundance in this work,
  particularly in the comparison in Fig.\,\ref{fig:gamma3}, may
  indicate that either the $\Gamma_{\rm e}$ in Sects.\,\ref{sec:fits},
  and \ref{sec:Mwind} are underestimated, due to the assumption of
  homogeneity, or the mass-loss predictions by \citet{gra1:08} are
  still too low. In any case, we do not expect a qualitative impact on
  our results, as the differential stellar properties within the
  sample are hardly affected.}

\subsection{The mass-loss properties of very massive stars}
\label{sec:disc_mdot}

The main result of the present study is the empirical confirmation of
a $\Gamma$-dependent mass-loss relation for stars approaching the
Eddington limit. This result is consistent with theoretical
predictions \citep{gra1:06,vin1:06,gra1:08,vin1:11}, as summarized in
Sect.\,\ref{sec:pred}.  A similar $\Gamma$-dependent relation has also
been proposed for the winds of LBVs, by \citet{vin1:02}.  When we
express our empirical results in the form $\dot{M}(\Gamma_{\rm e},
T_\star, L, X_{\rm H}^s)$, we find that the dependencies on $L$, and
$X_{\rm H}^s$ are very weak, indicating that $\Gamma_{\rm e}$, and
potentially $T_\star$, are the physically most relevant parameters.
{\changedC The temperature dependence in Eq.\,\ref{eq:mdotfit} is in
  remarkable} agreement with \citet{gra1:08}, however, {\changedC
  based on the present data, this result seems not fully established}.

The empirical confirmation of a $\Gamma$-dependence marks a paradigm
shift in the way we think about the development of stellar mass loss
through subsequent evolutionary phases. The physical effect that
causes (LBV and) WR-type mass loss is the proximity to the Eddington
limit. During their evolution, the cores of massive stars become
chemically enriched, and their $L/M$ ratio increases.  When Eddington
factors of order unity are reached, a strong WR-type stellar wind will
develop naturally.  With the removal of the outer envelope, in
combination with mixing processes, a hydrogen deficiency will develop
at the stellar surface.  It is therefore not the hydrogen deficiency
that defines the onset of the WR phase, as is usually assumed in
stellar evolution calculations.  Rather, the WR-phase starts as soon
as Eddington factors of order unity are reached.

This effect will have strong impact on the evolution of massive stars,
in terms of conventional mass loss, but also in terms of loss of
angular momentum \citep[e.g.,][]{lan2:98}.  For fast rotating stars,
the Eddington parameter effectively increases near the equator, due to
centrifugal forces. {\changedC The effect of rotation on mass loss has
  been studied for OB stars, e.g.\ by
  \citet{fri1:86,pel1:00,cur1:04,cur1:05,mad1:07}.  Close to
  rotational breakup (i.e., close to the $\Omega$-limit), the latter
  two of these works find shallow wind solutions, for which the mass
  loss does {\em not} significantly exceed the mass loss without
  rotation \citep[only by a factor of two, see][]{mad1:07}.  For
  WR-type winds, the effect of rotation has not yet been investigated.
  The strong $\Gamma$-dependence however implies that the combined
  effect of rotation and high $\Gamma$ may still be efficient at high
  rotational speed (i.e., close to the $\Gamma\Omega$-limit).  This
  effect may be of paramount importance for fast rotating WR stars,
  like the progenitors of long GRBs \citep[][]{yoo1:05}. At the
  present stage this is however highly speculative.}

In addition to that, the continuous transition between the O/Of, and
WNh stages (cf.\,Fig.\,\ref{fig:gamma3}) has consequences for our
theoretical understanding of the winds of very massive stars. Notably,
this suggests that they have the same driving mechanism, namely
radiation pressure. Nevertheless, there are important differences
between O-star winds and WR-winds. While the former are expected to be
optically thin at the sonic point, the latter are optically thick,
i.e., the wind is initiated and already accelerated in part below the
surface.  The transition between these two regimes has been resolved
by the model computations of \citet{vin1:11} with their different
$\Gamma$-dependencies for the OB, and the WR regime
(Eqs.\,\ref{eq:vink1} \& \ref{eq:vink2}).  \citeauthor{vin1:11} find
that this transition occurs for models with a wind efficiency $\eta =
\dot{M} \varv_\infty / (L/c) \approx 1$, which roughly corresponds to
the point where the winds become optically thick.

The properties of optically thick, WR-type winds have been suggested
to be determined by the so-called Fe-opacity peaks in deep atmospheric
layers, close to the sonic point
\citep[e.g.,][]{pis1:95,heg1:96,sch1:96,nug1:02}.  This has been
confirmed by WR wind models of \citet{gra1:05,gra1:08}, who proposed
that the formation of optically thick winds is supported close to the
Eddington limit, because the density scale height in the deep
atmospheric layers increases. In contrast to this, OB star winds are
thought to be dominated by the outer wind physics.  Within the
classical theory of radiatively driven winds \citep{cas1:75}, their
wind properties are determined at a critical point with a much higher
wind speed, which is typically of the order of the escape speed. The
wind energy at this point is thus of the same order of magnitude as
the gravitational energy. Small changes in the (effective)
gravitational potential thus only have small influence on the mass
loss of OB stars, i.e., the $\Gamma$-dependence is expected to be
weak.  The fact that we find observational evidence for a strong
$\Gamma$-dependence of WR mass loss, in agreement with the model
predictions, shows that there is indeed a substantial difference in
the way the mass-loss rates of OB and WR stars are determined.

{\changedC A strong temperature dependence of WR-type mass loss would}
support the importance of the deep layers for the wind physics of WR
stars.  \citet{gra1:08} predict a similar dependence, with
$\dot{M}\propto T_\star^{-3.5}$.  They explain this dependence by the
necessity that the critical point of the equation of motion needs to
be located in a specific temperature regime, and thus at higher
optical depths for cooler stars. They also show that this temperature
dependence is in agreement with the observed line strengths of
galactic WNL stars \citep[cf.\ Fig.\,2 in][]{gra1:08}.
Quantitatively, the temperature dependence might be of major
importance because it leads into phases of extremely high mass loss,
potentially even to a smooth transition into the LBV phase.  We note,
however, that our empirical result arises from a comparison of the
hotter O/Of stars vs.\ the cooler WNh stars.  Particularly for the
low-luminosity end of the O star sample we concluded in
Sect.\,\ref{sec:Mwind} that they might have a systematically different
internal structure than the WNh stars.  Moreover, for the relatively
weak winds of these objects, the concept of $\Gamma$-dependent mass
loss might be questionable.  The fact that we actually find a relation
that includes these objects is thus rather surprising. Because of
these concerns it would be desirable to back up this result by
complementary studies of WNh stars in different temperature regimes.

{\changedA In Fig.\,\ref{fig:gamma3} we compare our empirical relation
  (Eq.\,\ref{eq:mdotfit}) with theoretical mass-loss predictions by
  \citet{gra1:08} for different metallicities $Z$.  For a high $Z$ of
  $2\,Z_\odot$, we find a good quantitative agreement between theory
  and observation. Qualitatively, the slope of the empirical
  $\Gamma$-dependence is very well reproduced by the models.  The same
  holds for the relation by \citet[][]{vin1:11}. Their exponent of
  3.99 in Eq.\,\ref{eq:vink2} matches the exponent in {\changedC
    Eq.\,\ref{eq:mdotfit}} very precisely.  The major difference
  between the model predictions in Sect.\,\ref{sec:pred} thus lies in
  the strong temperature dependence that is predicted by
  \citet{gra1:08}.}

Summing up our results, we expect that massive stars close to the
Eddington limit tend to form WR-type winds with distinct properties
from classical radiatively driven winds. In the WR regime we find
mass-loss relations of the form $\dot{M}(\Gamma_{\rm e}, T_\star, L,
X_{\rm H}^s)$ which fundamentally depend on the Eddington parameter
$\Gamma_{\rm e}$.  Our theoretical models predict 1) a strong
$\Gamma$-dependence, 2) a weak dependence on $L$, and $X_{\rm H}^s$,
and 3) a strong temperature dependence, according to
\citet[][]{gra1:08}. {\changedA Empirically}, point 1), and 2) are
convincingly confirmed in our present study. The important temperature
dependence 3), is in agreement with our present results, {\changedC
  however, demands for further empirical confirmation.}

\subsection{Wind masses for very massive stars}
\label{sec:mwind}

{\changedA In Sect.\,\ref{sec:Mwind} we used the mass-loss relations by
  \citet{gra1:08} to derive the present masses of the Arches cluster
  stars, based on their observed mass-loss rates. The potential to
  estimate stellar masses in this way is a very interesting aspect of
  this work, in particular because our mass estimates} are
fundamentally different from wind masses that have {\changedA
  previously} been obtained for OB stars \citep[e.g.,][]{kud1:92}.

{\changedA In Sect.\,\ref{sec:disc_mdot}, we already discussed that}
the dynamics of classical radiation-driven winds is dominated by the
outer part of the wind, while WR-type winds are strongly influenced by
deep atmospheric layers.  Classical OB stars thus show a well-defined
relation between their escape velocity at the stellar surface, and
their terminal wind velocity, that helps to determine their masses
\citep{kud1:00,prin1:98,lam1:95}. Only \citet{vin1:02} previously used
a $\Gamma$-dependent mass-loss relation to constrain the stellar mass
of the LBV AG Car, however, their models still depend on the outer
wind physics, and thus on the ratio $\varv_\infty/\varv_{\rm esc}$.

{ WR-type winds depend mostly on the inner wind physics.  We have seen
  that this leads to a steep $\Gamma$-dependence for stars that are
  close to the Eddington limit, i.e., we are capable to determine
  precise $\Gamma_{\rm e}$, without consideration of the terminal wind
  velocity. This method is very promising, because the masses of stars
  close to the Eddington limit are already well known, {\changedB
    simply} because $\Gamma_{\rm e}$ is of the order of one. The wind
  models thus only need to provide an `improved guess' of $\Gamma_{\rm
    e}$.  Moreover, observational uncertainties are of minor
  importance because of the steepness of the $\Gamma$-dependence.}
The main uncertainties arise from the models, chiefly from the
contribution of metal lines to the `physical' Eddington factor
$\Gamma(r)$. However, even if this contribution would be
systematically wrong, the models still provide reliable {\em mass
  ratios}, i.e., the mass-loss relation can be calibrated with
observations. In the present work we have performed a first step
towards such a calibration. It seems, however, that the remaining
uncertainties in the metallicity, and the distance/extinction towards
the Arches cluster demand for complementary studies in different
environments.

\subsection{Evolutionary status of the Arches cluster stars}
\label{sec:evolution}

As a by-product of our present study we obtain information about the
{\changedC internal structure, and thus about the} evolutionary status
of the WNh stars in the Arches cluster.  Most importantly, we find
direct evidence that the largest part of these objects is still in the
phase of core hydrogen burning. In Fig.\,\ref{fig:gamma2} we have
shown that, if the stars were in the phase of core He-burning, 9 of
the 13 WNh stars in the sample would have Eddington factors
$\Gamma_{\rm e} > 0.75$ (cf.\ also Table\,\ref{tab:stars}).  Taking
into account that $\Gamma(r) > \Gamma_{\rm e}$ (Eq.\,\ref{eq:gammar}),
this would imply that these stars are extremely close, or even above
the physical Eddington limit.  The reason for this are the high
hydrogen surface abundances of these stars (up to solar).

{\changedC The $M$--$L$ relations used in this work are based on the
  assumption of a simple, chemically homogeneous internal structure of
  the sample stars. In Sects.\,\ref{sec:fits} \& \ref{sec:Mwind} we
find indications that this assumption may indeed be fulfilled by most
of the WNh stars in our sample. However, due to uncertainties in the
observational data, and the models, we cannot draw firm conclusions at
this stage.}

E.g., we found in Sect.\,\ref{sec:Mwind} that the wind masses of the
H-deficient WNh stars are in good overall agreement with the expected
masses of chemically-homogeneous stars
(cf.\,Fig.\,\ref{fig:MLarches}). This result however depends on the
adopted metallicity in our mass-loss relation.  In
Sect.\,\ref{sec:fits}, we found evidence for chemical homogeneity,
based on mass ratios, which are much more certain (cf.\,
Sects.\,\ref{sec:uncert} \& \ref{sec:mwind}).  Figs.\,\ref{fig:gamma1}
\& \ref{fig:gamma2} show that, under the assumption of chemical
homogeneity, the {\em relative masses} of the sample stars adjust in a
way that we observe a smooth $\dot{M}(\Gamma_{\rm e})$ relation. Under
the assumption of core He-burning, i.e., the completely in-homogeneous
case, we obtain differences in $\dot{M}$ by a factor of $\sim$10 for
similar $\Gamma_{\rm e}$, particularly for H-deficient WNh stars. This
suggests that {\changedC most of} the H-deficient WNh stars in the
Arches cluster are close to homogeneity, {\changedC although a similar
  behaviour may also originate from a stellar sample where all stars
  have a similar degree of inhomogeneity.}

{\changedC A chemically homogeneous structure is expected for single,
  fast rotating massive stars, e.g., according to evolutionary models
  by, \citet[][]{mae1:87,yoo1:06,bro1:11}.}  Note, however, that fast
rotation is no mandatory condition to achieve a nearly homogeneous
stellar structure. Because the convective cores of very massive stars
are very large, their remaining radiative envelopes {\changedC can
  easily be stripped off by strong mass loss \citep[cf.,][]{yun1:08}}.
E.g., for the WNh stars in our sample we find mass-loss timescales of
$\tau_{\rm wind}=M/\dot{M}\approx1..6\,10^6{\rm yr}$, similar to their
evolutionary timescales of $\tau_{\rm nuc}=7.2\,10^{10}{\rm yr}
\cdot(M/M_\odot)/(L/L_\odot) \approx 3..5\,10^6{\rm yr}$. In addition,
other scenarios, like mergers due to frequent stellar collisions in
dense clusters \citep[e.g.,][]{port1:99} might be considered in this
context.

An argument against chemical homogeneity are the low surface
temperatures of the Arches cluster stars (30--35\,kK).  Because of
their high mean molecular weight, homogeneous stars are expected to be
very compact, and to show very high surface temperatures.
\citet{mar1:09} indeed found evidence for homogeneous evolution for an
extremely early WN3h star in the SMC, based on its high temperature
($T_\star = 65$\,kK). This star is clearly located to the left of the
main sequence, but has still 50\% hydrogen at its surface, in
agreement with evolutionary tracks for fast rotating quasi homogeneous
stars by \citet{mey1:05}. So why are the Arches cluster stars so much
cooler?

In principle, their temperatures may indicate that the Arches stars
are evolved main sequence stars, and not homogeneous.  However,
\citet{ish1:99} found that very massive homogeneous stars at high
metallicities can have inflated stellar envelopes.  This effect is
expected to occur for stars close to the Eddington limit, and to be
enhanced by fast rotation \citep{mae1:08}.  The inflation is caused by
the Fe opacity peak at temperatures around 160\,kK. The effect is thus
metallicity dependent. Note that the same opacity peak is expected to
drive the inner winds of early WR subtypes \citep{nug1:02,gra1:05}.
However, for late WR subtypes, like the Arches stars, the opacity peak
is located below the wind driving zone, and {\changedC may} thus affect
the stellar envelope. The existence of the envelope extension has been
questioned by \citet{pet1:06} because in their hydrodynamic stellar
structure models the effect was suppressed by mass loss.  It is thus
not clear whether the envelope inflation occurs in nature, but it is
in line with the observed low temperatures of many H-free WR stars
\citep[e.g.,][]{ham1:06}, and the tendency that late WR subtypes are
preferentially found in high metallicity environments
\citep[e.g.,][]{cro1:07}.

\section{Conclusions}
\label{sec:concl}

Based on a semi-empirical study of the most massive stars in the
Arches cluster we could confirm key properties of WR-type stellar
winds that have been predicted in our previous theoretical studies
\citep{gra1:08,vin1:11}. {\changedA We find that mass loss is enhanced
  close to the Eddington limit, with a dependence of the form
  $\dot{M}(\Gamma_{\rm e}, T_\star, L, X_{\rm H}^s, Z)$ that shows} 1)
a strong dependence on the Eddington factor $\Gamma_{\rm e}$, 2) a
weak dependence on the luminosity $L$, and the hydrogen surface
abundance $X_{\rm H}^s$, and 3) a strong increase of mass loss for
decreasing stellar temperatures $T_\star$.  Due to the properties of
the sample, the latter {\changedC is however not fully established.}
The qualitative agreement between models and observations suggests
that the mass-loss properties of WR stars are determined in analogy
with the models, by the influence of the Fe-opacity peaks in deep
atmospheric layers.

The strong $\Gamma$-dependence marks an important paradigm shift with
respect to previous, luminosity-dependent relations, that are based on
purely empirical studies \citep{nug1:00,ham1:06}. The latter are
widely used, e.g., in stellar evolution models, where the WR-phase is
usually identified by a hydrogen-deficient surface composition.
According to our work, the physically relevant parameter, which {\em
  causes} WR-type mass loss, is the proximity to the Eddington limit.
A hydrogen deficiency then occurs naturally as a {\em consequence} of
the strong mass loss, {\changedA potentially in combination with
  rotational mixing.}

As a by-product of our work we obtain information on the evolutionary
status of the sample objects. {\changedA We find strong evidence that
  the luminous WNh stars in the Arches cluster are not evolved
  objects, but {\changed mostly} very massive, hydrogen-burning stars,
  supporting the picture that the most massive stars are WNh stars in
  young stellar clusters {\changedB
    \citep[e.g.,][]{dek1:97,cro1:10}}.}

The mass loss in the WR phase is of paramount importance for the
evolution, and the death of massive stars, {\changedA as well as the
  formation of long GRBs}.  With our present work we have made a first
step towards a more physically motivated description of WR-type mass
loss, incorporating the dependence on the Eddington factor
$\Gamma_{\rm e}$. Other predicted dependencies, e.g.\ on the stellar
temperature, and the metallicity, demand for similar studies of large
stellar samples in different environments.  Combined observational and
theoretical programs, like the ongoing VLT-FLAMES Tarantula Survey
{\changedC \citep[][Bestenlehner et al.\ in prep.]{eva1:11,bes1:11}},
will give improved insight into exactly these questions in the near
future.

\appendix
\section{Mass-luminosity relations for a broad range of stellar
  masses}
\label{sec:appendix1}

\begin{table*}
  \caption{Mass-luminosity relations for chemically-homogeneous stars. \label{tab:coeff}}
  \centering
  \begin{tabular}{lllrrrrrrrrrr} \hline \hline
    \rule{0cm}{2.2ex}{\changedC No.} & {ev.\ ph.} & Eq. & mass range & $F_1$ & $F_2$ & $F_3$ & $F_4$ & $F_5$ & $F_6$ & $F_7$ & $F_8$ & $F_9$ \\
    \hline \rule{0cm}{2.2ex}%
     1& H-b.  & \ref{eq:MLH_TXT}   & 12--250\,$M_\odot$  &  2.875 & $-$3.966 & 2.496 &  2.652 & $-$0.310 & $-$0.511 \\
     2& H-b.  & \ref{eq:MLH_TXT}   & 2--100\,$M_\odot$   &  1.967 & $-$2.943 & 3.755 &  1.206 & $-$0.727 & $-$0.026 \\
     3& H-b.  & \ref{eq:MLH_TXT}   & 60--4000\,$M_\odot$ &  3.862 & $-$2.486 & 1.527 &  1.247 & $-$0.076 & $-$0.183 \\
     4& H-b.  & \ref{eq:GAMH_TXT}  & 2--30\,$M_\odot$    & $-$2.688 & $-$7.843 & 2.471 &  2.758 & $-$0.233 & $-$0.747 \\
     5& H-b.  & \ref{eq:GAMH_TXT}  & 12--4000\,$M_\odot$ & $-$2.416 & $-$5.118 & 1.869 & $-$0.400 &  0.064 &  0.050 \\
     6& He-b. & \ref{eq:MLHE_TXT}  & {\changedA 8}--250\,$M_\odot$  &  3.017 &  2.446 & $-$0.306 \\
     7& He-b. & \ref{eq:MLHE_TXT}  & 0.6--100\,$M_\odot$ &  2.635 &  2.986 & $-$0.488 \\
     8& He-b. & \ref{eq:MLHE_TXT}  & 60--1000\,$M_\odot$ &  3.826 &  1.619 & $-$0.099 \\
     9& He-b. & \ref{eq:GAMHE_TXT} & 0.3--100\,$M_\odot$ & $-$2.204 &  1.831 &  0.149 \\
    10& He-b. & \ref{eq:GAMHE_TXT} & 12--500\,$M_\odot$  & $-$1.676 &  1.075 &  0.404 \\
    \hline \rule{0cm}{2.2ex}%
    11& H-b.  & \ref{eq:LMH_TXT1}, \ref{eq:LMH_TXT2} & 12--250\,$M_\odot$  &  4.026 & 4.277 & $-$1.0 & 25.48 &  36.93 & $-$2.792 & $-$3.226 & $-$5.317 & 1.648 \\
    12& H-b.  & \ref{eq:LMH_TXT1}, \ref{eq:LMH_TXT2} & 2--100\,$M_\odot$   &  2.582 & 0.829 & $-$1.0 & 9.375 &  0.333 &  0.543 & $-$1.376 & $-$0.049 & 0.036 \\
    13& H-b.  & \ref{eq:LMH_TXT1}, \ref{eq:LMH_TXT2} & 60--4000\,$M_\odot$ &  10.05 & 8.204 & $-$1.0 & 151.7 &  254.5 & $-$11.46 & $-$13.16 & $-$31.68 & 2.408 \\
    14& H-b.  & \ref{eq:GMH_TXT1}, \ref{eq:GMH_TXT2} & 2--30\,$M_\odot$    &  5.303 & 5.918 & $-$1.0 & 16.58 & $-$4.292 & $-$72.89 & $-$7.881 & $-$13.76 & 3.206 \\
    15& H-b.  & \ref{eq:GMH_TXT1}, \ref{eq:GMH_TXT2} & 12--4000\,$M_\odot$ & $-$14.60 & 3.125 &  1.0 & 251.0 &  15.63 &  72.24 &  18.20 &  12.21 & 0.781 \\
    16& He-b. & \ref{eq:LMHE_TXT} & {\changedA 8}--250\,$M_\odot$  &  3.997 &         $-$1.0 & 25.83 & $-$3.268 \\
    17& He-b. & \ref{eq:LMHE_TXT} & 0.6--100\,$M_\odot$ &  3.059 &         $-$1.0 & 14.76 & $-$2.049 \\
    18& He-b. & \ref{eq:LMHE_TXT} & 60--1000\,$M_\odot$ &  8.177 &         $-$1.0 & 105.5 & $-$10.10 \\
    19& He-b. & \ref{eq:GMHE_TXT} & 0.3--100\,$M_\odot$ & $-$6.144 &          1.0 & 52.54 &  6.711 \\
    20& He-b. & \ref{eq:GMHE_TXT} & 12--500\,$M_\odot$  & $-$1.330 &          1.0 & 5.919 &  2.475 \\
    \hline \end{tabular}
  \tablefoot{
    The table describes the mass-luminosity relations 
    $L(M,X_{\rm H})$, and $M(L, X_{\rm H})$ that are derived in this work, as well as 
    the relations $\Gamma_{\rm e}(M, X_{\rm H})$,
    and $M(\Gamma_{\rm e}, X_{\rm H})$ in the appendix.
    For each relation we give the assumed evolutionary phase, the corresponding equation numbers, the relevant mass range,
    and the numeric coefficients $F_n$. }
\end{table*} 

\begin{figure}[tp]
  \parbox[b]{0.49\textwidth}{\includegraphics[scale=0.435]{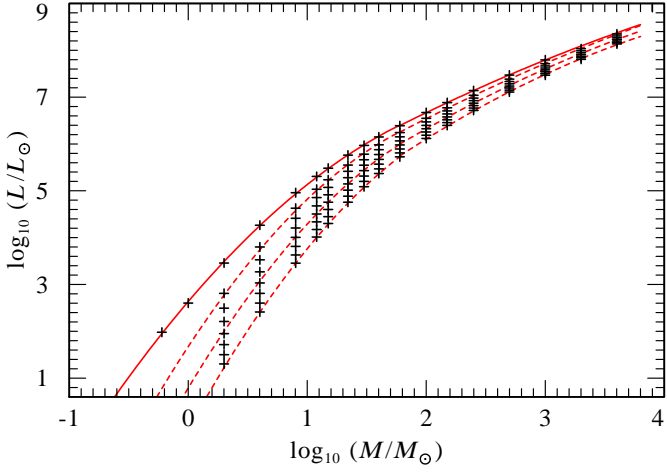}}
  \caption{Homogeneous stellar structure models for the mass range
    0.6--4000\,$M_\odot$ (black symbols). The models are computed for
    hydrogen mass fractions $X_{\rm H} = 0.0$--0.7, in steps of 0.1
    (from top to bottom).  Dashed red lines indicate the $M$-$L$
    relations for $X_{\rm H} = 0.1$, 0.4, and 0.7 according to
    Eq.\,\ref{eq:MLH_TXT}\,\&\,Table\,\ref{tab:coeff}, and the solid
    red line corresponds to pure He models
    (Eq.\,\ref{eq:MLHE_TXT}\,\&\,Table\,\ref{tab:coeff}).}
  \label{fig:MLapp}
\end{figure}

\begin{figure}[tp]
  \parbox[b]{0.49\textwidth}{\includegraphics[scale=0.435]{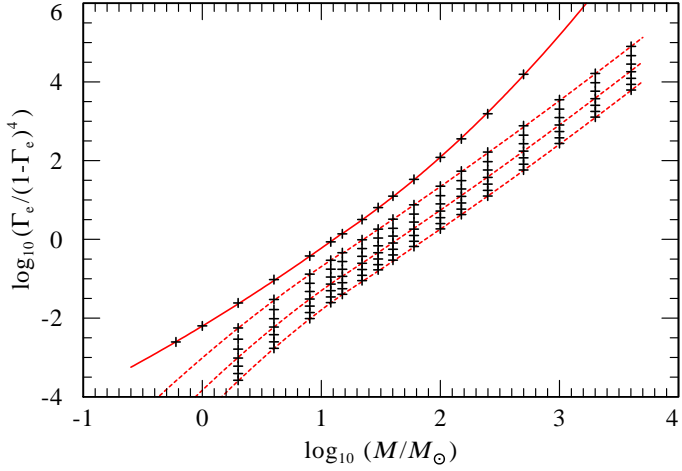}}
  \caption{Eddington parameters for the same models as in
    Fig.\,\ref{fig:MLapp}. The quantity $G_4 = \Gamma_{\rm
      e}/(1-\Gamma_{\rm e})^4$ is theoretically expected to follow a
    relation with $G_4\propto M^2$ (see Eq.\,\ref{eq:G4}). The symbols
    in this plot are the same as in Fig.\,\ref{fig:MLapp}.}
  \label{fig:G4app}
\end{figure}

In Sect.\,\ref{sec:masses} of this work we provided fitting formulae
for mass-luminosity relations $L(M, X_{\rm H})$, and the corresponding
masses $M_{\rm hom}(L, X_{\rm H})$, and $M_{\rm Heb}(L)$ for massive,
chemically-homogeneous stars. We focused on a mass range of
12--250\,$M_\odot$, which covers the observed properties of the most
massive stars known.  Within this mass range we obtained a very good
fit quality for $L(M,X_{\rm H})$, with a fitting error of up to
0.02\,dex.  The resulting fitting relations are however strictly
limited to this mass range. The reason is that just within this range
the $M$--$L$ relation changes from a power law $L\propto M^\eta$, with
$\eta \approx 3.8$ for low masses, to a nearly linear relation with
$\eta \rightarrow 1$ for very massive stars. Our fits in
Sect.\,\ref{sec:masses} describe this `bending' of the $M$--$L$
relation very well, but are not capable {\changedB of reproducing} the
asymptotic behaviour for higher and lower masses.  We {\changedB
  therefore} decided to provide separate fitting relations in this
appendix, that cover the range of 2--4000\,$M_\odot$ for H-burning
stars, and 0.6--1000\,$M_\odot$ for He-burning stars.

In analogy to Sect.\,\ref{sec:masses}, we obtain separate fits to our
computational results for H-burning stars with $X_{\rm H} = 0.1$--0.7,
and He-burning stars ($X_{\rm H}=0$). The resulting relations have the
same form as Eqs.\,\ref{eq:MLH_TXT}\,\&\,\ref{eq:MLHE_TXT} in
Sect.\,\ref{sec:masses}. Moreover, we obtain separate relations for
the steep low-mass part, and the flat high-mass part of the $M$--$L$
relation.  The corresponding coefficients are given in
Table\,\ref{tab:coeff}.  The maximum fitting errors in
$\log(L/L_\odot)$ are 0.05 for H-burning stars with masses of
2--100\,$M_\odot$, 0.02 for H-burning stars with 60--4000\,$M_\odot$,
0.03 for He-burning stars with 0.6--100\,$M_\odot$, and 0.003 for
He-burning stars with 60--1000\,$M_\odot$.  The results are displayed
in Fig.\,\ref{fig:MLapp}.

Inverting these relations, we obtain $M_{\rm hom}(L, X_{\rm H})$, and
$M_{\rm Heb}(L)$ in the same way as in Sect.\,\ref{sec:masses}, i.e.,
$M_{\rm hom}(L, X_{\rm H})$ is given by
Eqs.\,\ref{eq:LMH_TXT1}\,\&\,\ref{eq:LMH_TXT2}, and $M_{\rm Heb}(L)$
by Eq.\,\ref{eq:LMHE_TXT}, with coefficients from
Table\,\ref{tab:coeff}.

\section{Asymptotic behaviour of the Eddington factor for large
  stellar masses}
\label{sec:appendix2}

For the highest stellar masses, the $M$--$L$ relation {\changedB has
  to become nearly linear, because $\Gamma_{\rm e}\rightarrow 1$.} As
$\Gamma_{\rm e}$ itself hardly changes in this range, the value of
$1-\Gamma_{\rm e}$ becomes physically more relevant. Our previous
fitting relations, however, chiefly focus on $M$, and $L$, i.e., on
the correct reproduction of $\Gamma_{\rm e}$.  The resulting value of
$1-\Gamma_{\rm e}$ can be significantly affected by small
uncertainties due to fitting errors.  E.g., for the largest masses of
up to $4000\,M_\odot$, our relation {\changedC in
  Eq.\,\ref{eq:MLH_TXT}} reproduces the stellar luminosity and thus
$\Gamma_{\rm e}$ rather precisely, but the error in $1-\Gamma_{\rm e}$
amounts to 0.1\,dex. To improve on this we take advantage of a
theoretically predicted scaling relation that has recently been
pointed out by \citet{owo1:11}, and goes back to the original work by
\citet{edd1:18}.

{\changedA With Eddington's original definition
  \begin{equation}
\label{eq:GammaP}
(1-\beta) \equiv \frac{P_{\rm rad}}{P} = \frac{a T^4}{3 P},
\end{equation}
the total pressure $P$ due to gas + radiation can be written in the
form
\begin{equation}
\label{eq:P}
P = P_{\rm gas} + P_{\rm rad} = \frac{{\cal R}}{\mu}\rho T + \frac{a}{3}T^4
= \left( \frac{3{\cal R}^4}{a \mu^4}\right)^\frac{1}{3}
\left( \frac{(1-\beta)}{\beta^4}\right)^\frac{1}{3} \rho^\frac{4}{3}.
\end{equation}
Assuming $\beta$} is a constant, this relation leads to a polytropic
equation of state with $P\propto \rho^{4/3}$, which is the basis of
the Eddington Standard Model.

{\changedA Eddington's assumption that $\beta$ is a constant, holds
  very well in large parts of the stellar envelope where the energy
  transport is dominated by radiation, and where the opacity is
  dominated by the constant electron scattering opacity $\chi_{\rm
    e}$. In that case it follows from the equation of radiative
  diffusion that
\begin{equation}
  \frac{\partial P_{\rm rad}}{\partial r} =
  \frac{\chi_{\rm e} \rho L}{4\pi r^2 c} =
  \Gamma_{\rm e} \frac{G{\changedC M(r)}\,\rho}{r^2} =
  \Gamma_{\rm e}\frac{\partial P}{\partial r}.
\label{eq:XXX}
\end{equation}
The latter equality follows from the definition of the classical
Eddington factor $\Gamma_{\rm e}$ (Eq.\,\ref{eq:gammae}), and the
equation of hydrostatic equilibrium.  {\changedC Under the assumption
  that $\Gamma_{\rm e}$ is a constant, Eq.\ref{eq:XXX} is just the
  derivative of the relation $P_{\rm rad}=\Gamma_{\rm e}P$, i.e., it
  becomes equivalent to Eq.\,\ref{eq:GammaP}. We thus have} $P_{\rm
  rad}/P = (1-\beta) = \Gamma_{\rm e}$, and with Eq.\ref{eq:P},
\begin{equation}
  \label{eq:P3}
  P^3 = \left( \frac{3{\cal R}^4}{a \mu^4}\right)
  \left( \frac{\Gamma_{\rm e}}{(1-\Gamma_{\rm e})^4}\right) \rho^4.
\end{equation}

}

Here we are interested in how {\changedA $\Gamma_{\rm e}$} changes
qualitatively with the stellar mass $M$. For this purpose we express
$P$, and $\rho$ in {\changedA Eq.\,\ref{eq:P3}} by two-point scaling
relations in $M$, and $R$. From hydrostatic equilibrium it follows
directly that $P/R \approx G M \rho / R^2$.  With $\rho \approx M/R^3$
this leads to $P \approx M^2/R^4$.  Using the last two relations with
{\changedA Eq.\,\ref{eq:P3}}, we obtain the approximate scaling
relation
\begin{equation}
\label{eq:G4}
\frac{\Gamma_{\changedA \rm e}}{(1-\Gamma_{\changedA \rm e})^4} \propto \frac{P^3}{\rho^4} \propto M^2.
\end{equation}

Based on this simple scaling law we obtain very precise fitting
relations for the quantity
\begin{equation}
\label{eq:G4def}
G_4 \equiv \frac{\Gamma_{\rm e}}{(1-\Gamma_{\rm e})^4}.
\end{equation}
The resulting fits to our computational results are plotted in
Fig.\,\ref{fig:G4app}.  For H-burning, chemically-homogeneous stars
with $X_{\rm H} = 0.1$--0.7 we use a relation of the form
\begin{eqnarray}
\label{eq:GAMH_TXT}
\log(G_4) & = & [F_1 + F_2\,\log(1+X_{\rm H})]\\
\nonumber 
& + & [F_3 + F_4\,\log(1+X_{\rm H})]\,\log(M/M_\odot) \\
\nonumber
& + & [F_5 + F_6\,\log(1+X_{\rm H})]\,\log(M/M_\odot)^2.
\end{eqnarray}
In Table\,\ref{tab:coeff} we give the coefficients $F_1$--$F_6$, for
the mass ranges 2--30\,$M_\odot$, and 12--4000\,$M_\odot$.  For pure
He stars ($X_{\rm H} =0$) we use
\begin{eqnarray}
\label{eq:GAMHE_TXT}
\log(G_4) & = & F_1 + F_2\,\log(M) + F_3\,\log(M)^2,
\end{eqnarray}
with coefficients $F_1$--$F_3$ from Table\,\ref{tab:coeff}, for the
mass ranges of 0.3--100\,$M_\odot$, and 12--500\,$M_\odot$.  The
maximum fitting error in $\log(G_4)$ for relations (\ref{eq:GAMH_TXT}
\& \ref{eq:GAMHE_TXT}) is 0.03.

For given $G_4$, $\Gamma_{\rm e}$ can be obtained from the solution of
Eq.\,\ref{eq:G4def}

\begin{equation}
  \Gamma_{\rm e} = 1 - \frac{C}{D^{1/6}}\sqrt{\frac{E}{G_4}}\,
  \left(\sqrt{\frac{3\sqrt{D\,G_4}}{C\,E^{3/2}}-1} - 1\right),
\end{equation}
with the definitions
\begin{eqnarray}
  C & = & \sqrt{1/12},\\
  D & = & (C/3\sqrt{256\,G_4 + 27} + 1/2)/G_4^2,\\
  E & = & 3\,G_4\,D^{2/3}-4.
\end{eqnarray}

Finally, we obtain relations for the corresponding masses $M_{\rm
  hom}(G_4, X_{\rm H})$, and $M_{\rm Heb}(G_4)$ from
Eqs.\,\ref{eq:GAMH_TXT} \& \ref{eq:GAMHE_TXT}.
\begin{equation}
\label{eq:GMH_TXT1}
\log(M_{\rm hom}/M_\odot) = \frac{F_1 + F_2\,\log(1+X_{\rm H})
                     + F_3 \sqrt{g}}{1 + F_9\,\log(1+X_{\rm H})},
\end{equation}
with
\begin{equation}
\label{eq:GMH_TXT2}
\begin{array}{l}
  g = F_4 + F_5 \log(G_4) + F_6 \log(1+X_{\rm H})^2   \\
  \rule{0cm}{2.4ex}\hspace{0.975cm} + (F_7 + F_8 \log(G_4))\log(1+X_{\rm H})
\end{array}
\end{equation}
and
\begin{equation}
\label{eq:GMHE_TXT}
\log(M_{\rm Heb}/M_\odot) = F_1 + F_2\, \sqrt{F_3 + F_4 \log(G_4)}.
\end{equation}
The coefficients $F_n$ are given in Table\,\ref{tab:coeff}.

\bibliographystyle{aa} \bibliography{aamnem99,astro}


\end{document}